\documentclass[preprint2]{aastex} 
\usepackage[T1]{fontenc}
\usepackage{verbatim}
\usepackage{graphicx}
\usepackage{amssymb}
\usepackage[unicode=true, pdfusetitle, bookmarks=true,bookmarksnumbered=false,bookmarksopen=false,  breaklinks=false,pdfborder={0 0 1},backref=false,colorlinks=false]{hyperref}
\usepackage{natbib}

\makeatletter
\slugcomment{}
\shorttitle{}
\shortauthors{}

\makeatother

\begin{document}


\title{Structure and Magnetic Fields in the Precessing Jet System SS\,433\\III. Evolution of the Intrinsic Brightness of the Jets from a Deep Multi-Epoch VLA Campaign}

\author{Michael R. Bell,\altaffilmark{1} David H. Roberts, and John F. C. Wardle}
\affil{Department of Physics, MS-057, Brandeis University\\ 415 South Street, Waltham, MA 02453}
\email{\href{mailto:mrbell@mpa-garching.mpg.de}{mrbell@mpa-garching.mpg.de},
\href{mailto:roberts@brandeis.edu}{roberts@brandeis.edu}, \href{mailto:wardle@brandeis.edu}{wardle@brandeis.edu}}

\altaffiltext{1}{Currently at the Max Planck Institute for Astrophysics, Karl-Schwarzschildstrasse 1, 85741 Garching, Germany.}

\begin{abstract}
We present a sequence of five deep observations of SS\,433 made over the summer of 2007 using the VLA in the A configuration at 5 and 8 GHz. In this paper we study the brightness profiles of the jets and their time evolution. We also examine the spectral index distribution in the source. We find (as previously reported from the analysis of a single earlier image) that the profiles of the east and west jets are remarkably similar if projection and Doppler beaming are taken into account. The sequence of five images allows us to disentangle the evolution of brightness of individual pieces of jet from the variations of jet power originating at the core. We find that the brightness of each piece of the jet fades as an exponential function of age (or distance from the core), $e^{-\tau/\tau'}$, where $\tau$ is the age at emission and $\tau' = 55.9 \pm 1.7$ days. This evolutionary model describes both the east and west jets equally well.  There is also significant variation (by a factor of at least five) in jet power with birth epoch, with the east and west jets varying in synchrony. The lack of deceleration between the scale of the optical Balmer line emission ($10^{15}$ cm) and that of the radio emission ($10^{17}$ cm) requires that the jet material is much denser than its surroundings. We find that the density ratio must exceed 300:1.
\end{abstract}




\section{Introduction}\label{sec:introduction}

%
%
%

SS\,433 was the first known microquasar, a type of x-ray binary (XRB) system consisting of a collapsed star accreting material from a less evolved donor star. The feature that distinguishes a microquasar from other XRBs is that they are variable radio sources that can eject pairs of oppositely directed relativistic jets \citep{mirabel99}. In many ways they are analogous to their AGN counterparts only smaller in size resulting in much shorter variability time scales.  This provides a distinct observational advantage for studying accreting black holes and relativistic jet systems. 

SS\,433 has been intensely studied since the discovery of oscillating simultaneously red and blue shifted hydrogen lines in the optical spectrum \citep{margon79a}. A kinematic model was proposed that explained these features as the result of a pair of mildly-relativistic precessing jets \citep{margon79b}. This model was soon confirmed and several parameters disambiguated by comparison with early VLA radio images \citep{hjellming81}. The jets precess with a period of 163 days about a cone with a semi-opening angle of $\sim 20^\circ$. The central axis is inclined from the line of sight by $\sim 80^\circ$. For in-depth reviews of previous work, see \citet{margon84} and \citet{fabrika06}. 

Previously in \cite{roberts08} we analyzed the structure of SS\,433 in both total and polarized intensity at $0.1\arcsec$ resolution from a 15 GHz VLA image. In \citet{roberts10} (henceforth Paper~II) we described in detail the procedure for measuring the \textit{intrinsic brightness} of the jets, a measure of the source emission as observed in a comoving reference frame, and analyzed the profile of a single, high dynamic range observation.  We found that the east and west jet profiles were consistent with an assumption of intrinsic symmetry and that the profile was more complex than could be fit by a simple exponential or power law. We suggested that the profile shape may be indicative of variation in the power injected into the jet.

In this paper we present a series of deep two-frequency VLA observations of the jets of SS\,433 and use them to study the jets as they evolve over an 80 day period.  Proper characterization of the jet dynamics is important for testing and developing jet models which predict, amongst other things, the brightness decay rate based on assumed physical parameters.  

In Section \ref{sec:observations} we describe the observations and data reduction and present the images.  In Section \ref{sec:spectral_index} we present maps of the spectral index across the source.  In Section \ref{sec:intrinsic_brightness} we calculate the intrinsic brightness profile at each epoch and investigate how it evolves in time. We discuss the implications of our results in Section \ref{sec:discussion} and present our conclusions in Section \ref{sec:conclusions}.  

\section{Observations}\label{sec:observations}

In order to study the evolution of the arcsecond scale radio emission from SS\,433, we conducted a series of five VLA A-array observations made over the summer of 2007.\footnote{VLA project code AR0637} Each observation includes data at two frequencies: 5 GHz ($\lambda6$~cm, C-band) and 8.5 GHz ($\lambda3.6$~cm, X-band). Table \ref{tab:obs_overview} provides a summary of the observing dates and the precessional phases for the observations. 

Frequent observations of the nearby, unresolved calibrator source J1950+081 were included for phase calibration purposes. Two flux calibrator sources, 3C286 and 3C48, were observed at the beginning and end of the observations, respectively. Additionally, a 5 minute scan of the bright source 3C84 ($\sim$14 Jy at 5 GHz as of 2007) was included to allow for the calibration of baseline-dependent complex gains. 

These observations were deemed ``shared-risk'' because the VLA was in the midst of being upgraded to the EVLA. This meant that the array was in a transitional state, with ten antennas having received new hardware while the others were still equipped with legacy hardware. The upgraded antennas had been equipped with new electronics, including the IF/LO system, receivers, and bandpass filters. Also, on June 27th, the ``MODCOMP'' computer system that was responsible for controlling the array was upgraded to the EVLA monitor and control system. The WIDAR correlator was not yet in use at this time.  The number of VLA and EVLA antennas used in each observation, after data flagging, is shown in Table \ref{tab:obs_overview}.


All data were calibrated using the NRAO Astronomical Image Processing Software (AIPS) package. We followed the standard procedure outlined in the AIPS Cookbook \citep{AIPS}, modified to include baseline calibration. The baseline calibration procedure is not typically necessary when processing VLA data, but due to the mismatched bandpass filters between VLA and EVLA antennas, this step was essential to reduce closure errors. Imaging and self-calibration were carried out using DIFMAP \citep{shepherd94}.

Due to last minute re-scheduling and problems with the array, the July 1st data at both frequencies and the August 24th data at 5 GHz did not include observations of a flux calibrator. Instead, the flux of the phase calibrator was determined from the other observations and used to set the absolute flux scale. The flux of J1950+081 was roughly constant throughout the summer, measuring $1.09\pm0.02$ Jy at C-band and $0.78\pm0.02$ Jy at X-band. We estimate the uncertainty in our flux scales as $\pm 3\%$.

\subsection{The Images}\label{ssec:images}

\begin{figure}
	\centering
	\plottwo{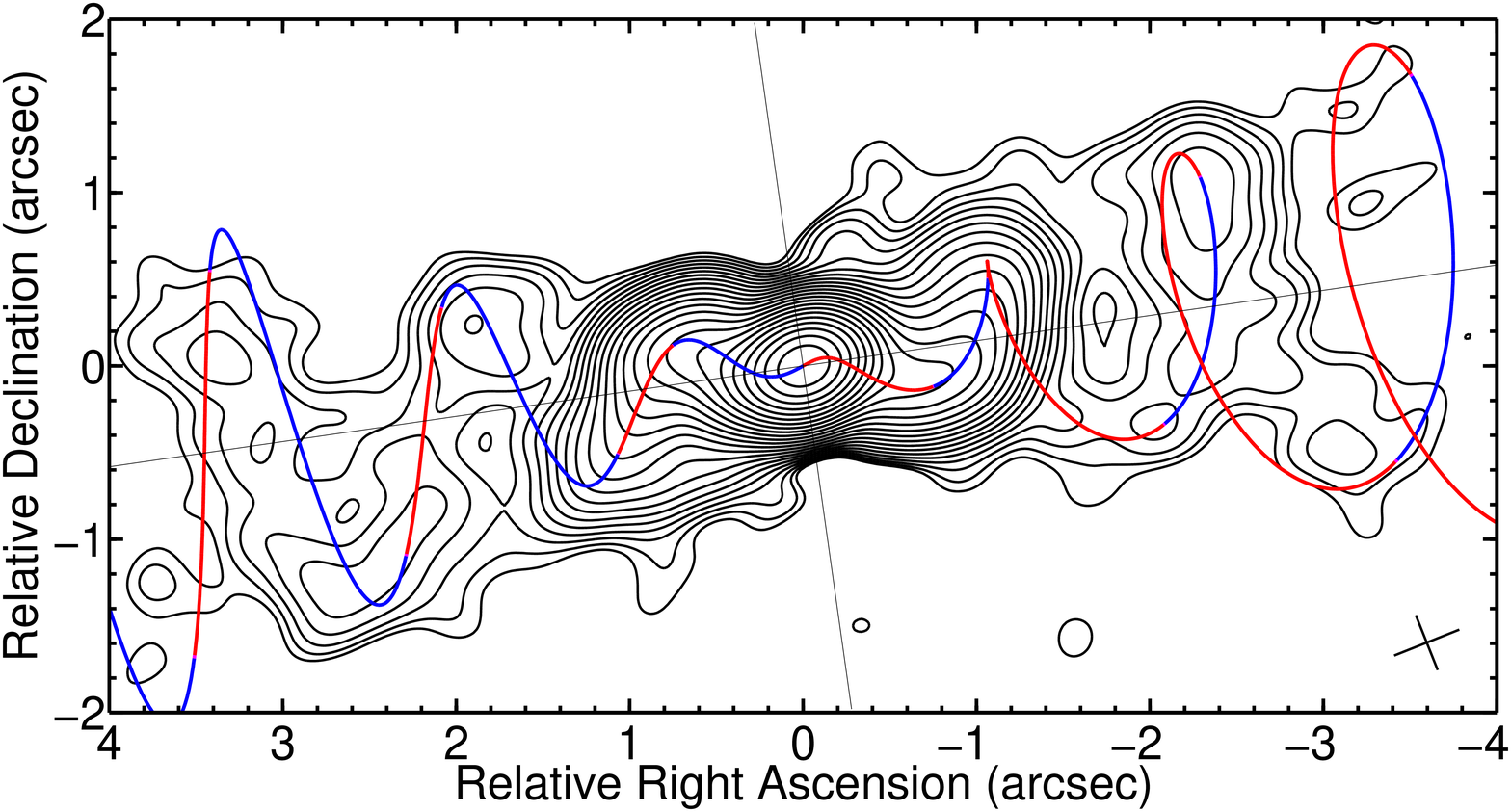}{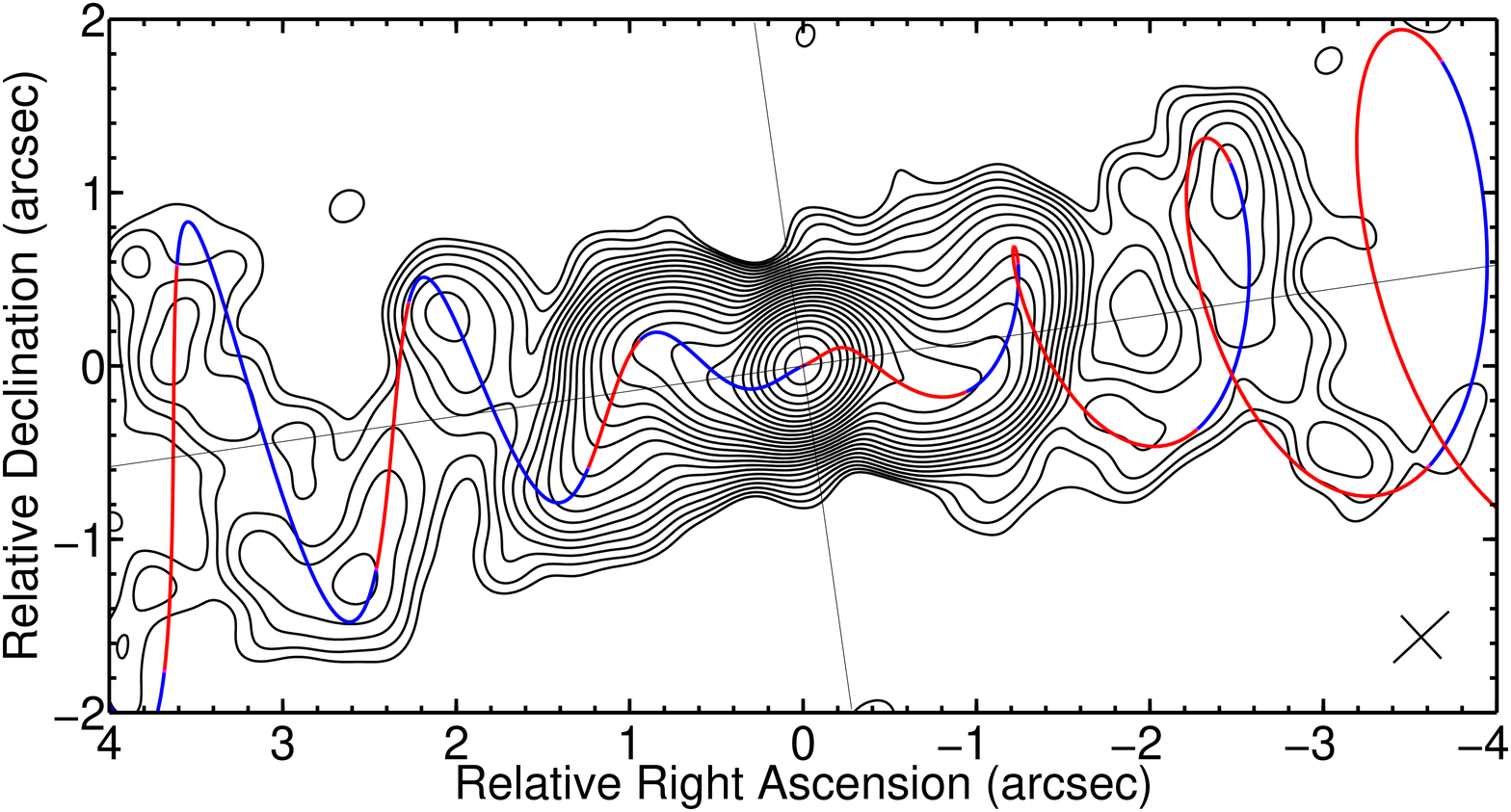}
	\plottwo{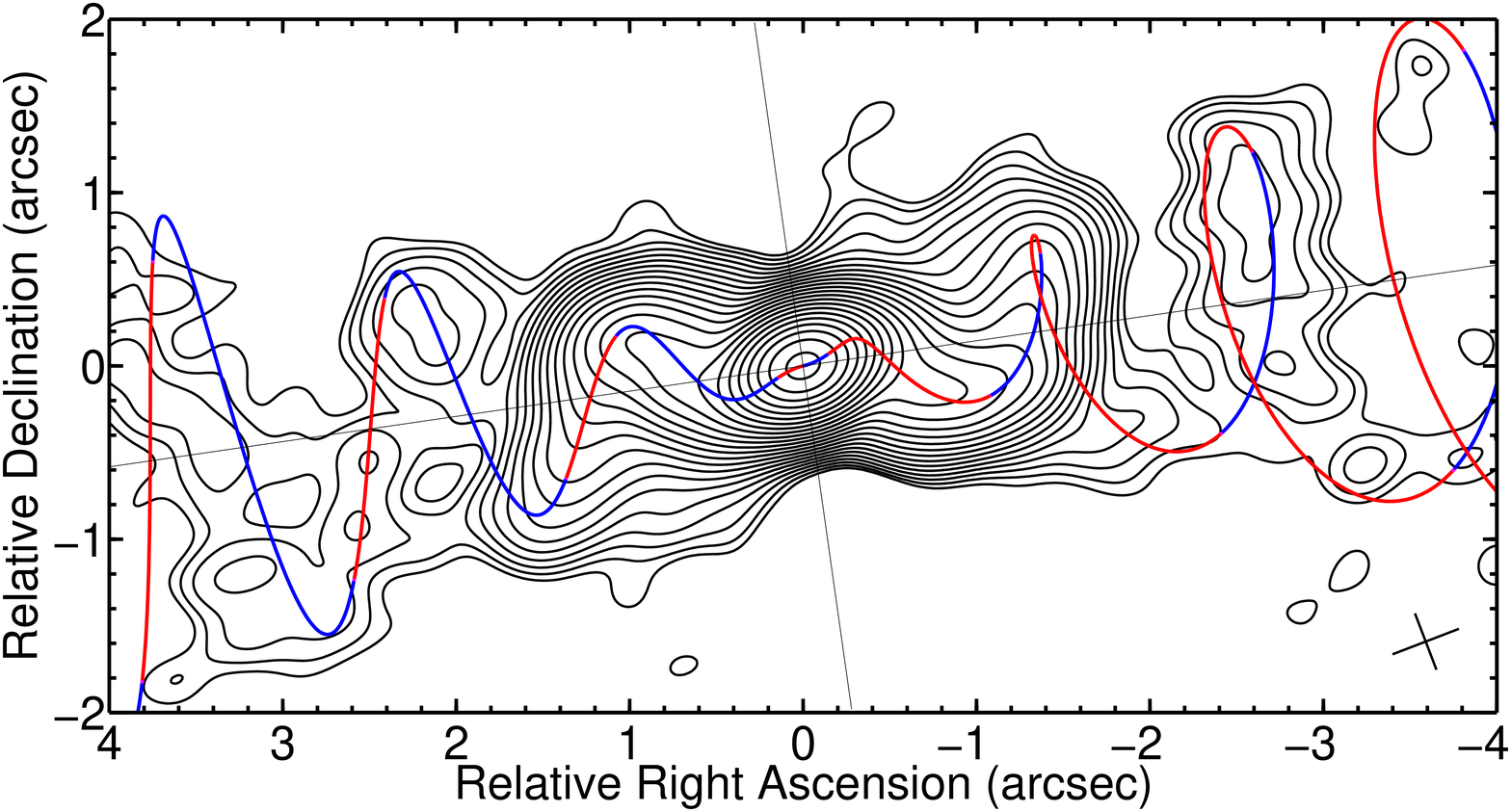}{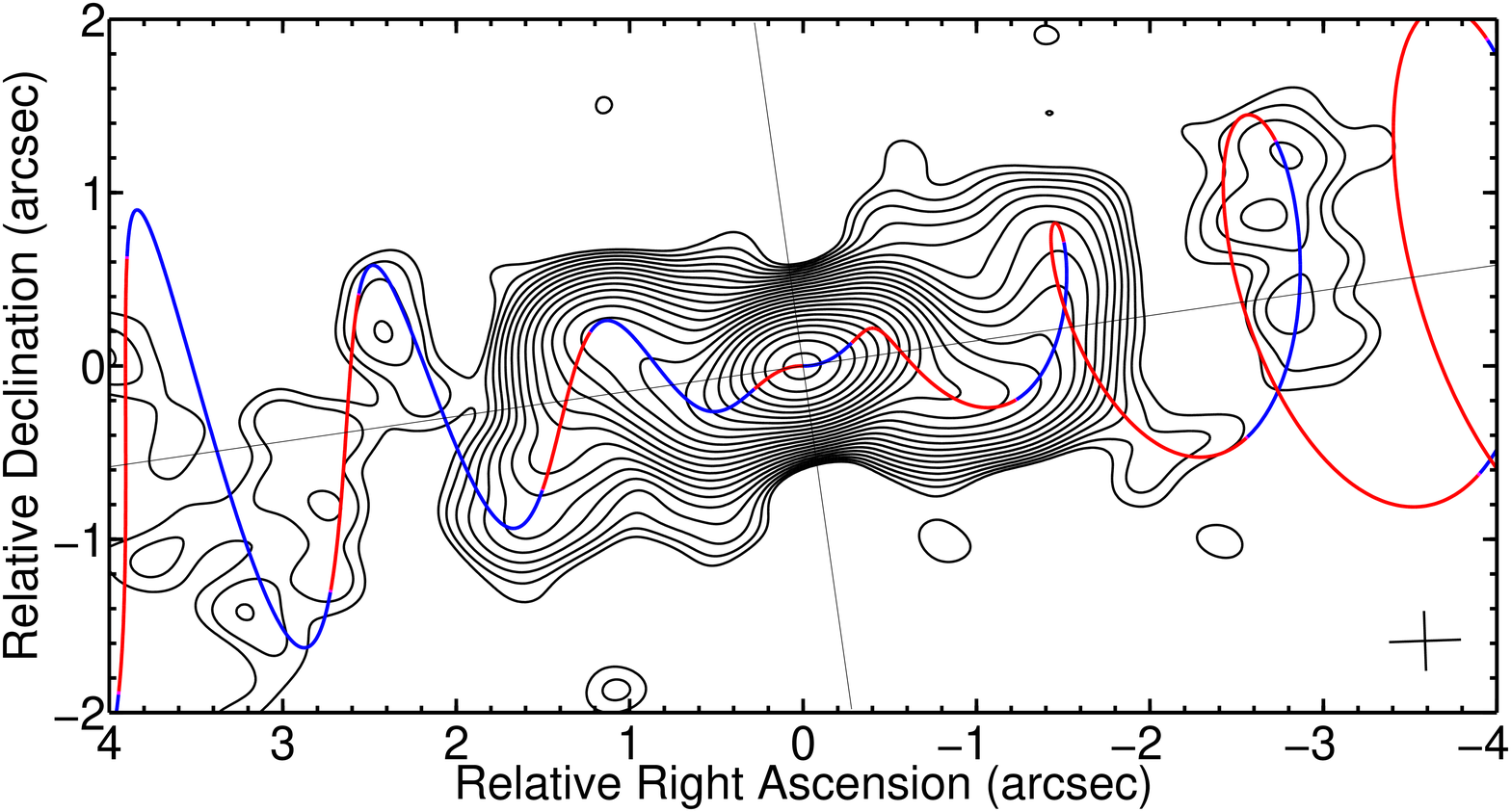}
	\epsscale{0.5}\plotone{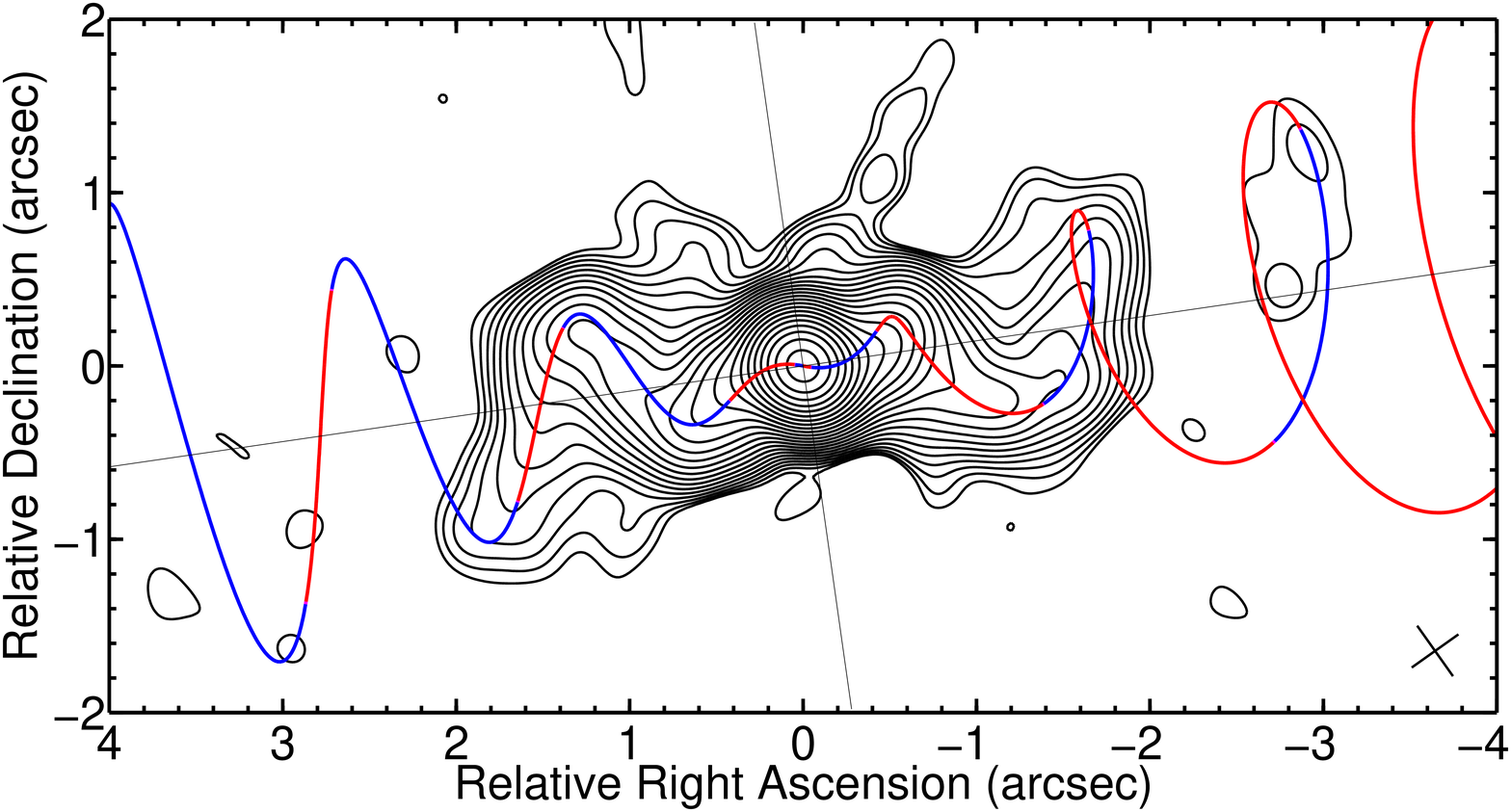}\epsscale{1.0}
	\caption[C-band Images]{C-band uniformly weighted images of total intensity. Contours are in steps of $\sqrt{2}$. The restoring beam is represented by a cross in the lower right of each image. Overlayed on each figure is the kinematic model with blue and red line segments representing the retreating and oncoming jet material, respectively. Detailed information about contour levels, map noise, etc. is found in Table \ref{tab:c_image_details}. (a) 2007 June 8, (b) 2007 July 1, (c) 2007 July 18, (d) 2007 August 5, (e) 2007 August 24.}
	\label{fig:cuni}
\end{figure}

\begin{figure}
	\centering
	\plottwo{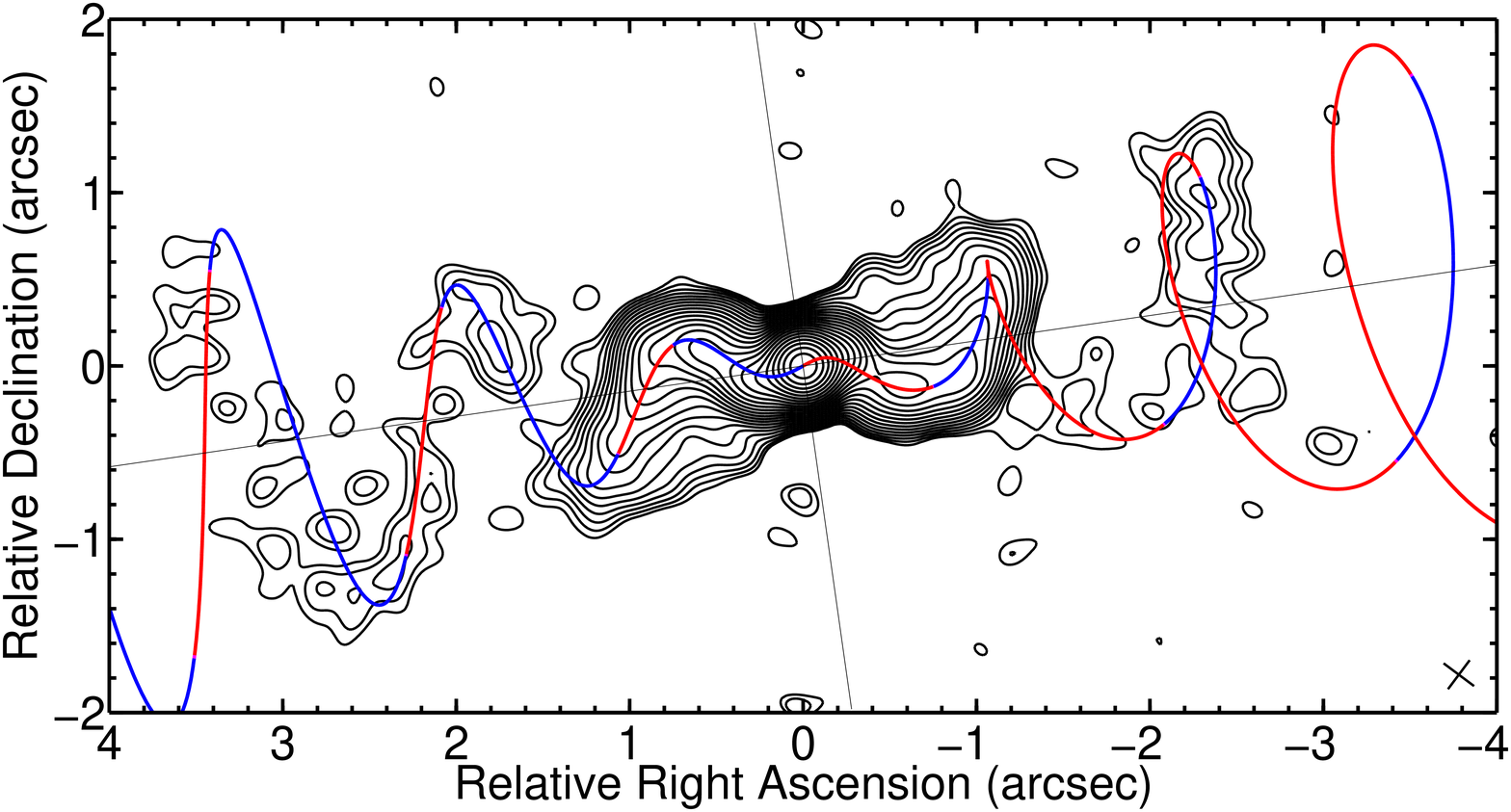}{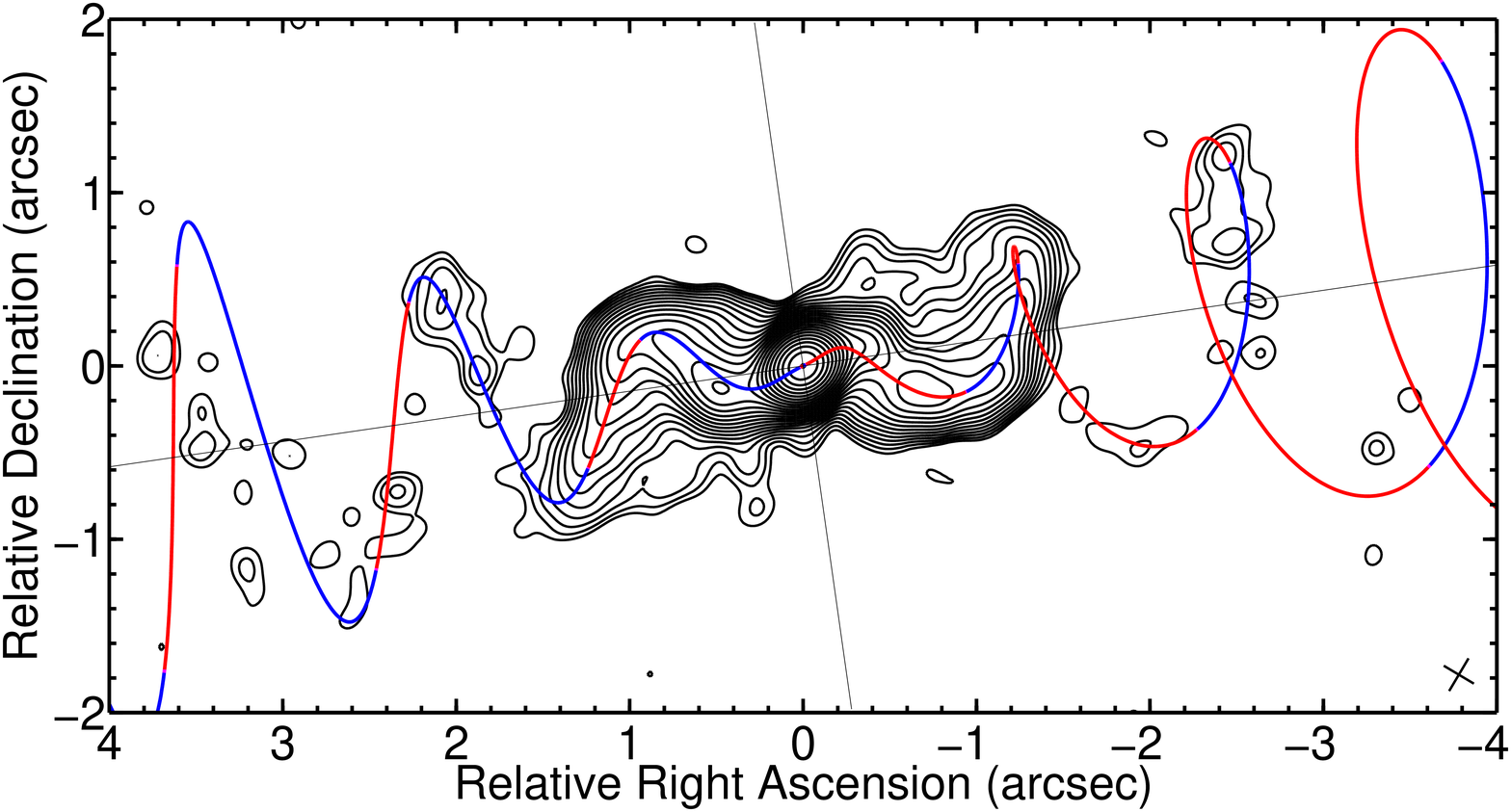}
	\plottwo{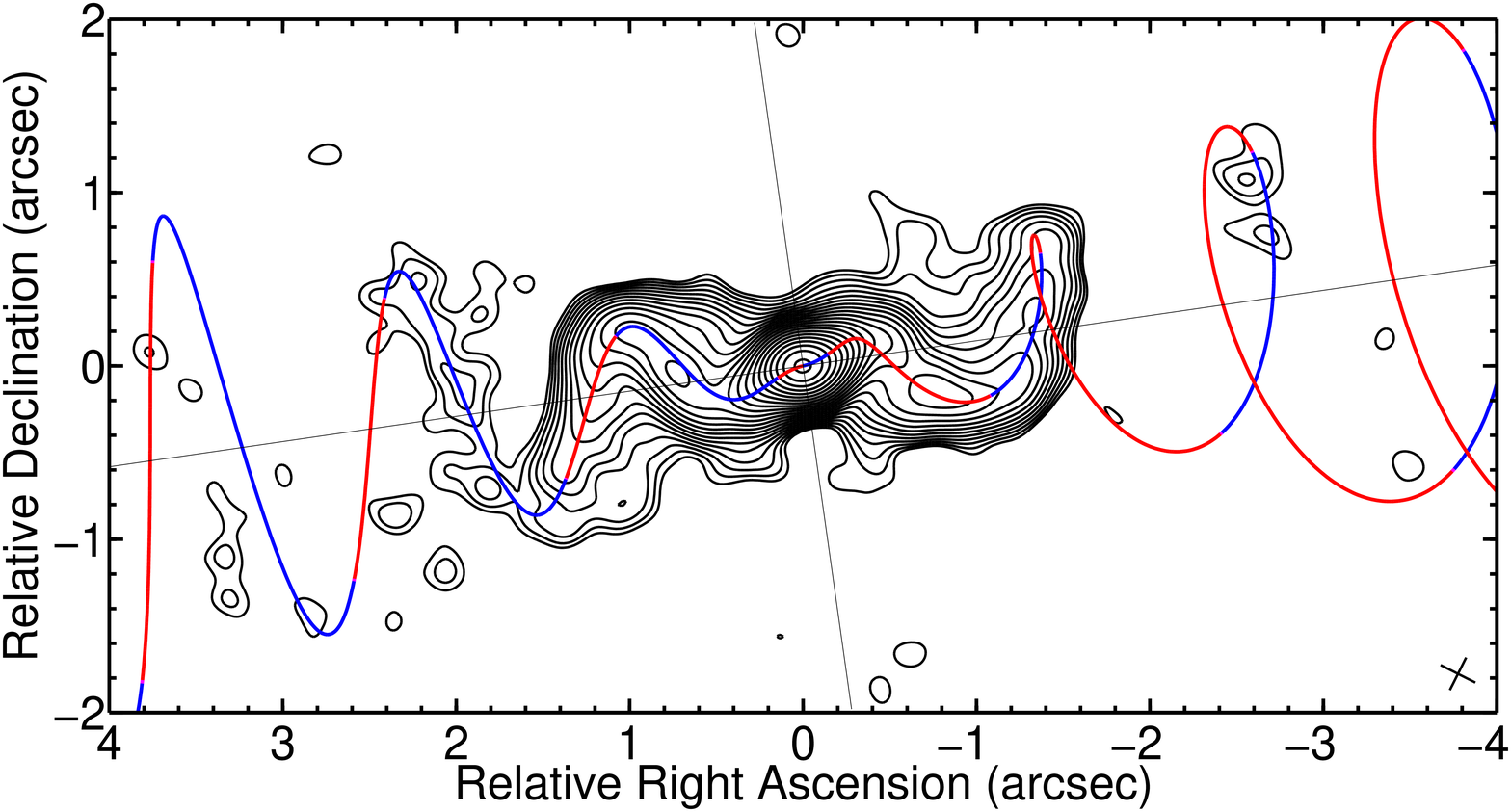}{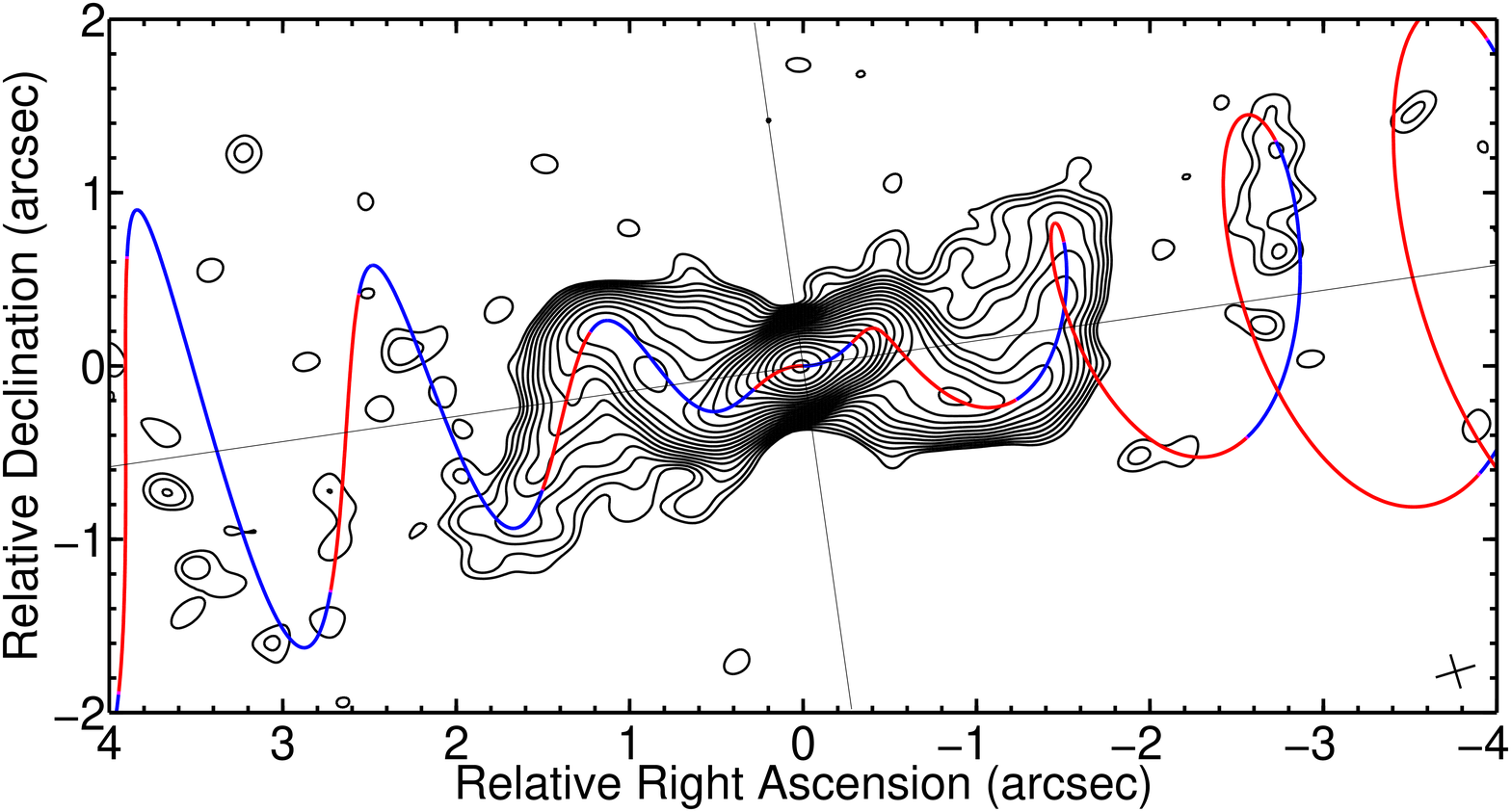}
	\epsscale{0.5}\plotone{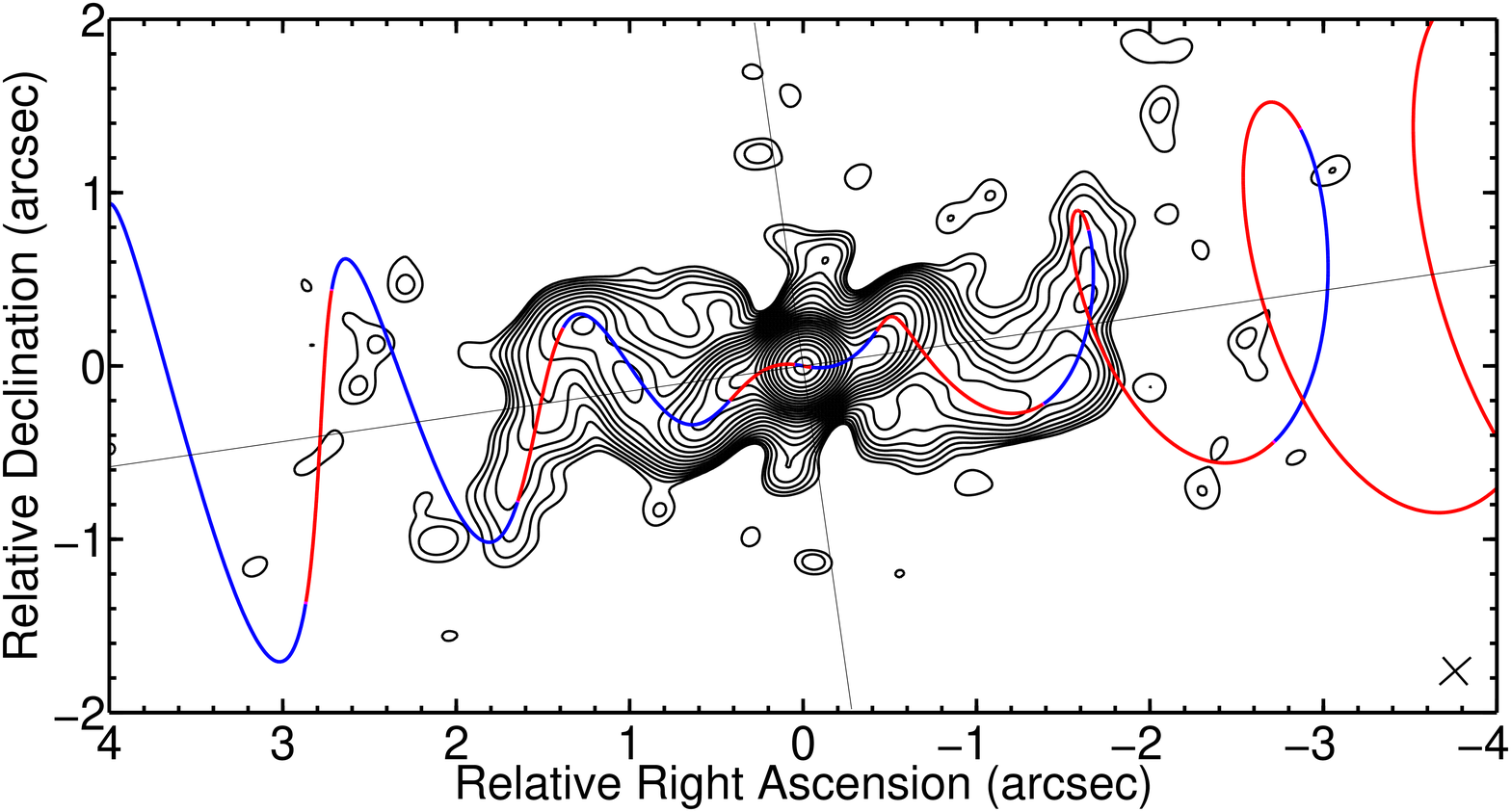}\epsscale{1.0}
	\caption[X-band Images]{X-band uniformly weighted images of total intensity. Contours are in steps of $\sqrt{2}$. The restoring beam is represented by a cross in the lower right of each image. Overlayed on each figure is the kinematic model with blue and red line segments representing the retreating and oncoming jet material, respectively. Detailed information about contour levels, map noise, etc. is found in Table \ref{tab:x_image_details}. (a) 2007 June 8, (b) 2007 July , (c) 2007 July 18, (d) 2007 August 5, (e) 2007 August 24.}
	\label{fig:xuni}
\end{figure}

Figures \ref{fig:cuni} and \ref{fig:xuni} show the uniformly weighted total intensity images at each epoch for 5 and 8.5 GHz, respectively. The plots are of total intensity with a step size of $\sqrt{2}$ between contours. The CLEAN beam is shown as a cross in the lower right hand corner. More information about each image can be found in Tables \ref{tab:c_image_details} and \ref{tab:x_image_details}. In these tables the columns marked ``BMAJ'' and ``BMIN'' provide the FWHM angular dimensions of the major and minor axes of the restoring beam; the beam position angle is listed in the column marked ``BPA''.  The $I_{max}$ and $I_{min}$ values are the map peak and bottom contour intensities, respectively.  The ``RMS'' value is the average of the root mean squared intensity measured in four empty regions of the image.  For some images the RMS has also been calculated by fitting a histogram of the pixel intensity values to a Gaussian.  Both methods yield the same results.  The thermal noise limit of the VLA for these observations as calculated using the VLA exposure calculator\footnote{\href{http://www.vla.nrao.edu/astro/guides/exposure/}{http://www.vla.nrao.edu/astro/guides/exposure/}} is approximately $20 \mu$Jy for the C-band images and $14\mu$Jy for the X-band images. The RMS noise level achieved in each image is typically 2-3 times these values. 

The simple kinematic model is overlaid on each image with blue and red line segments indicating material that is traveling toward and away from the observer, respectively. We adopt the ephemeris of \citet{eikenberry01} and assume a distance to the source of 5.5 kpc \citep{blundell04}.  A summary of the model parameters may be found in Table 1 of Paper~II. We include neither the nutation \citep{katz82} nor orbital velocity variation \citep{blundell07} since they imply structure on a smaller scale than the $\sim350\mbox{ mas}$ resolution of our images. In Paper~II these perturbations to the basic model were included as a method of sampling the area around the mean model curve in order to estimate the degree to which our measurements varied by changing the model location. Here we use an alternate method for determining this uncertainty as described below.

\begin{figure}
	\centering
	\plotone{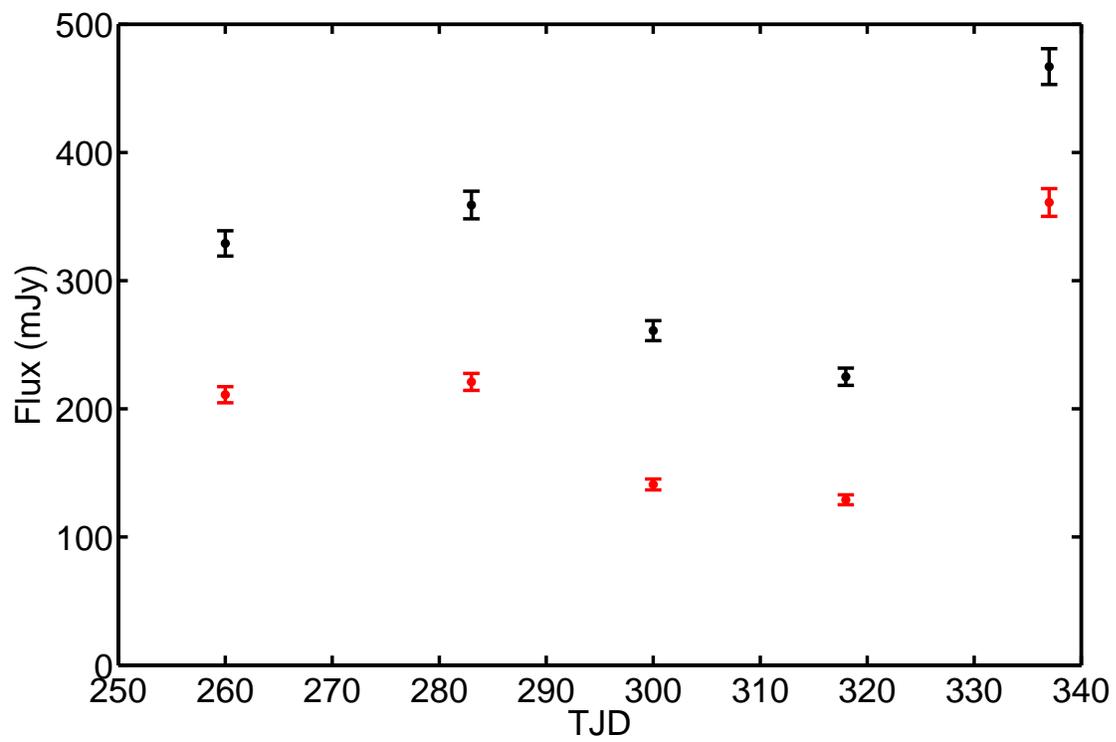}
	\caption[Core Variability]{The core flux as a function of TJD for (black) 5 GHz and (red) 8.5 GHz. The error bars represent a 3\% calibration uncertainty. The 5 and 8.5 GHz fluxes vary in the same manner, and each roughly doubles between 2007 August 5 and 2007 August 24.}
	\label{fig:core_v_tjd}
\end{figure}

The core flux in each observation is shown as a function of truncated Julian date (TJD = JD - 2454000) in Figure \ref{fig:core_v_tjd}. The error bars in the figure represent the $3\%$ calibration uncertainty. This is much larger than the uncertainty due to noise in the image and is a conservative estimate based on the stability of the measured flux of the phase calibrator. We find that the 5 and 8.5 GHz fluxes vary in the same fashion and that the flux more than doubles in the three weeks between the final two observations, increasing from $0.23$ Jy to $0.47$ Jy at 5 GHz. 

\section{The Spectral Index}\label{sec:spectral_index}

The resolution of the VLA at 5 and 8.5 GHz can be approximately matched by using the appropriate weighting schemes, thus allowing for pixel-by-pixel measurement of the spectral index $\alpha$ ($S_\nu \propto \nu^{-\alpha}$). It is important that we understand the extent to which the spectral index changes along the jet when we calculate the intrinsic brightness (see Section \ref{sec:intrinsic_brightness}) since the Doppler beaming correction depends on this quantity. 

To produce spectral index maps we use the uniformly weighted 5 GHz data and the naturally weighted 8.5 GHz data. The average beam diameter (at FWHM) of the uniformly weighted 5 GHz images is $370$ milliarcseconds (mas), while the average beam diameter for the naturally weighted 8.5 GHz images is $280$ mas. The CLEAN models for each frequency were convolved with a $350$ mas circular restoring beam so that the two images could be compared pixel-by-pixel. The spectral index is simply
\begin{equation}
	\alpha=-\frac{\ln S_{C}/S_{X}}{\ln\nu_{C}/\nu_{X}}.
\end{equation}

\begin{figure}
	\centering
	\plottwo{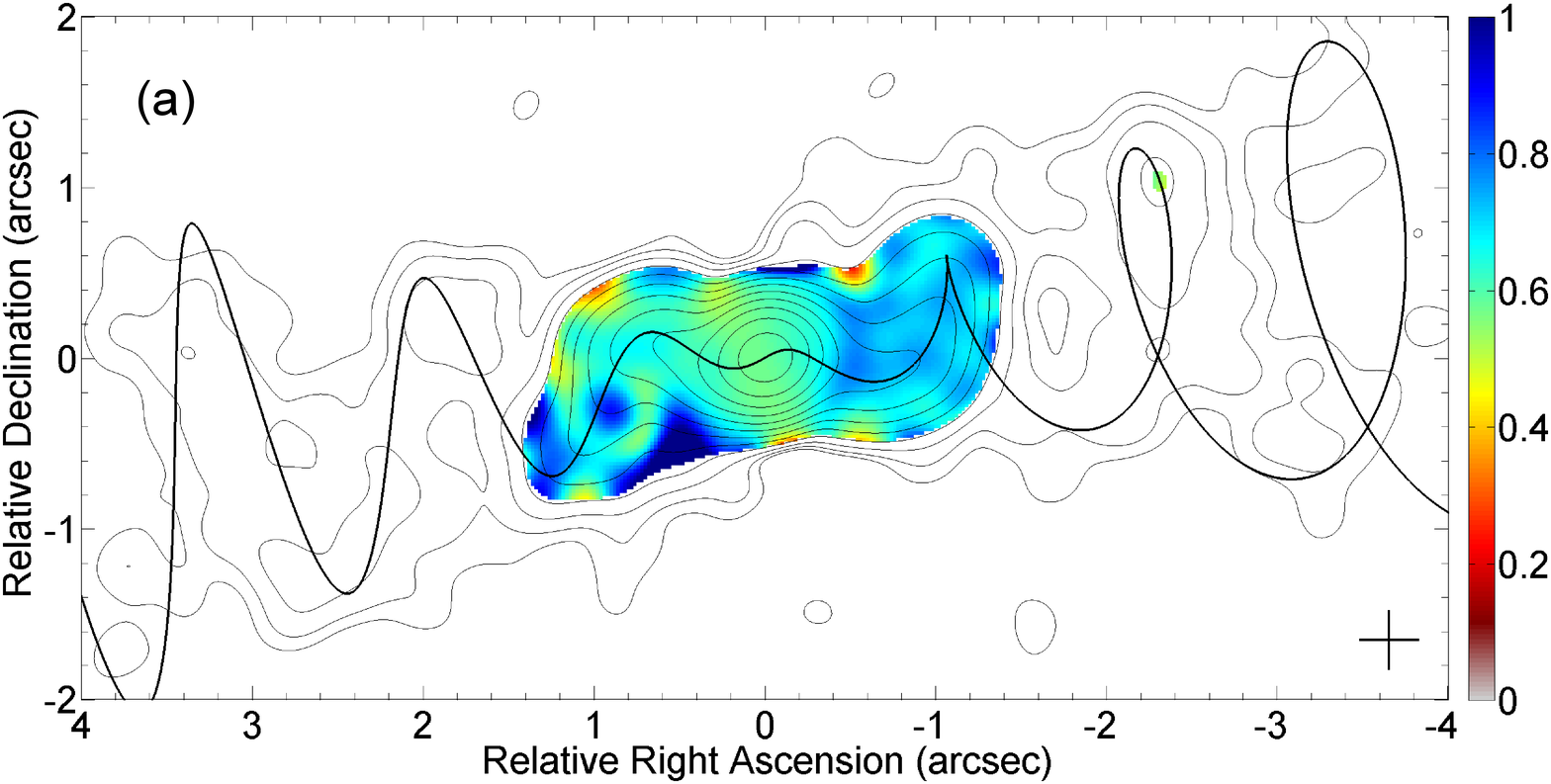}{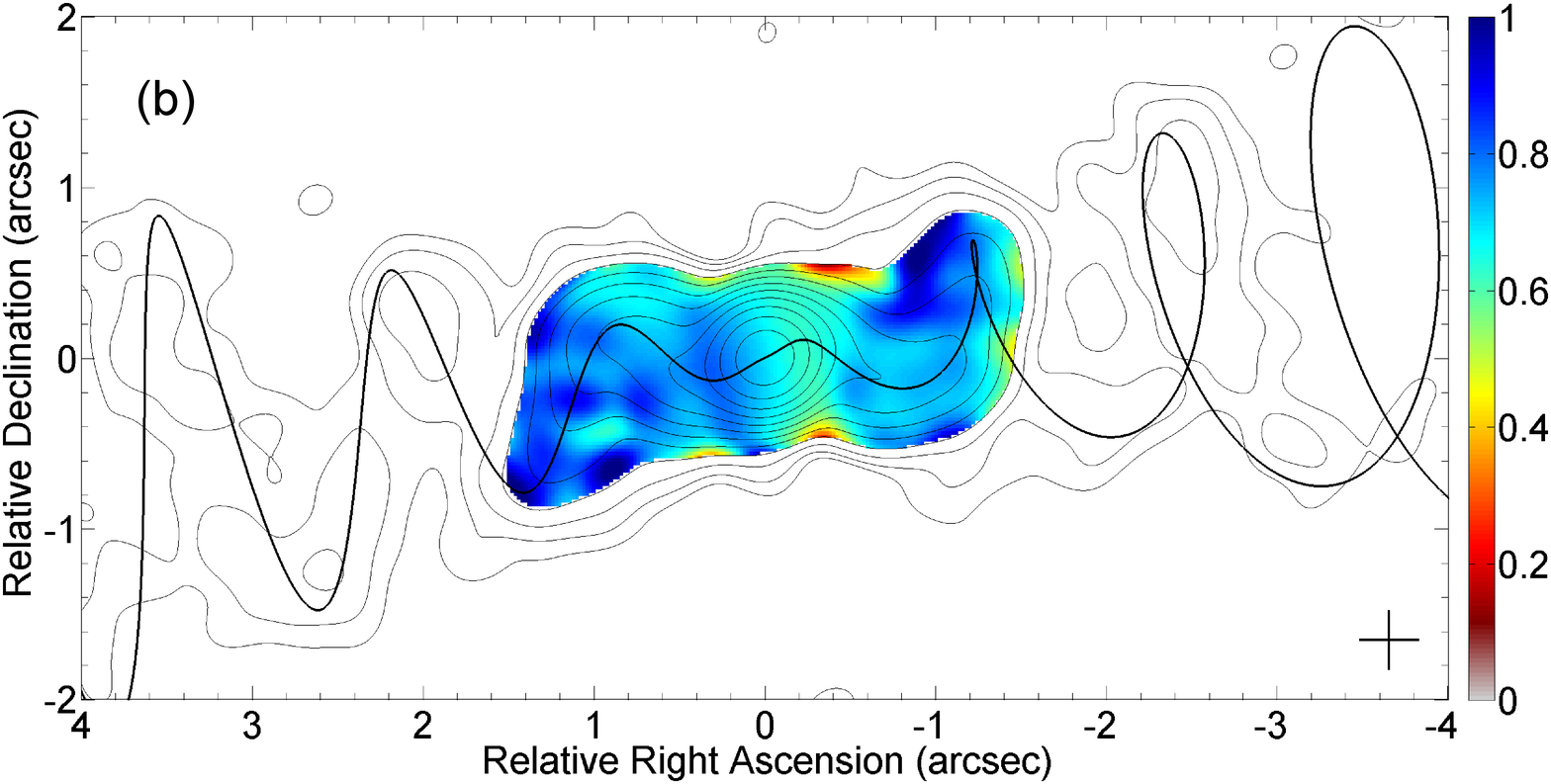}
	\plottwo{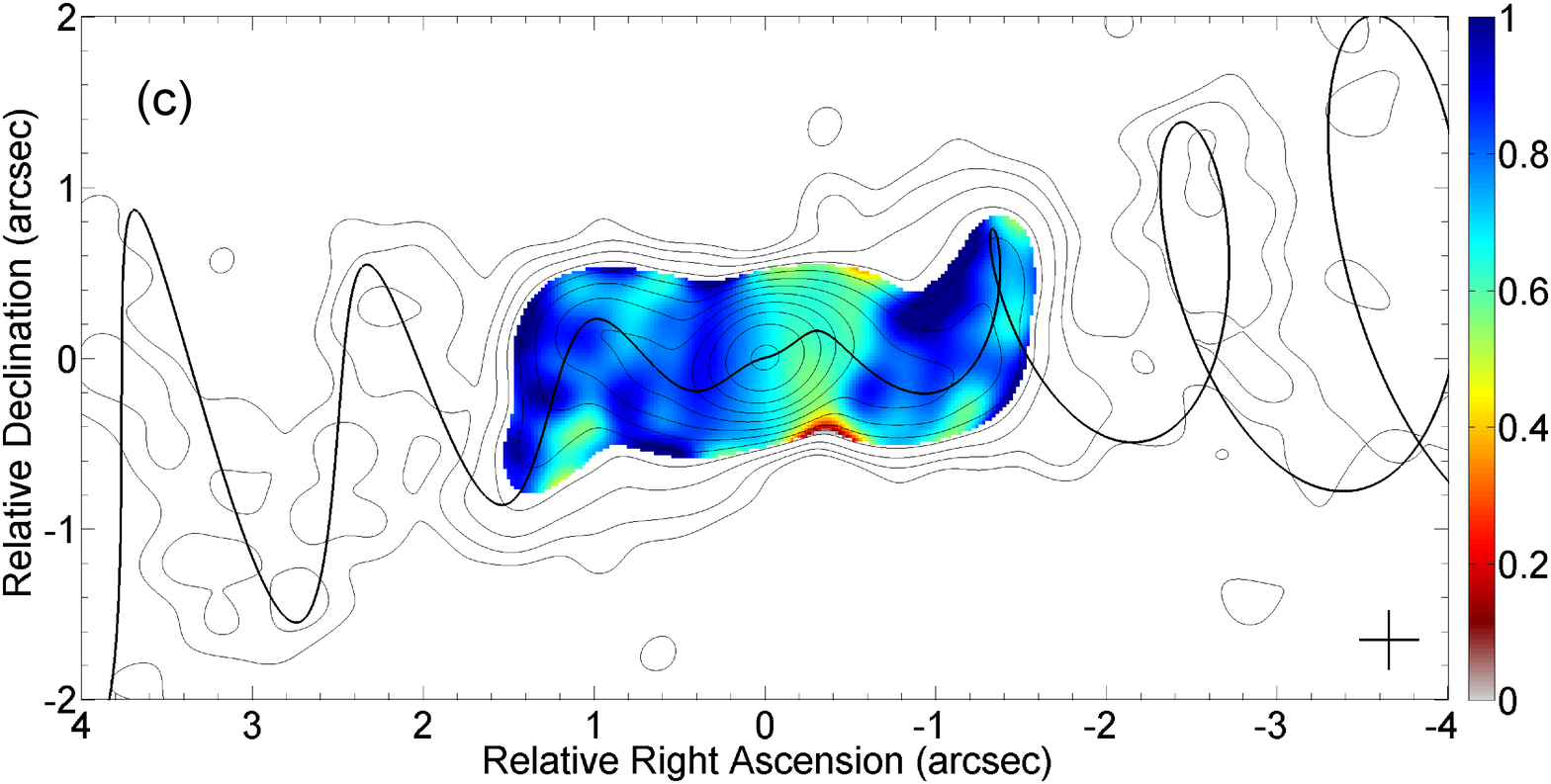}{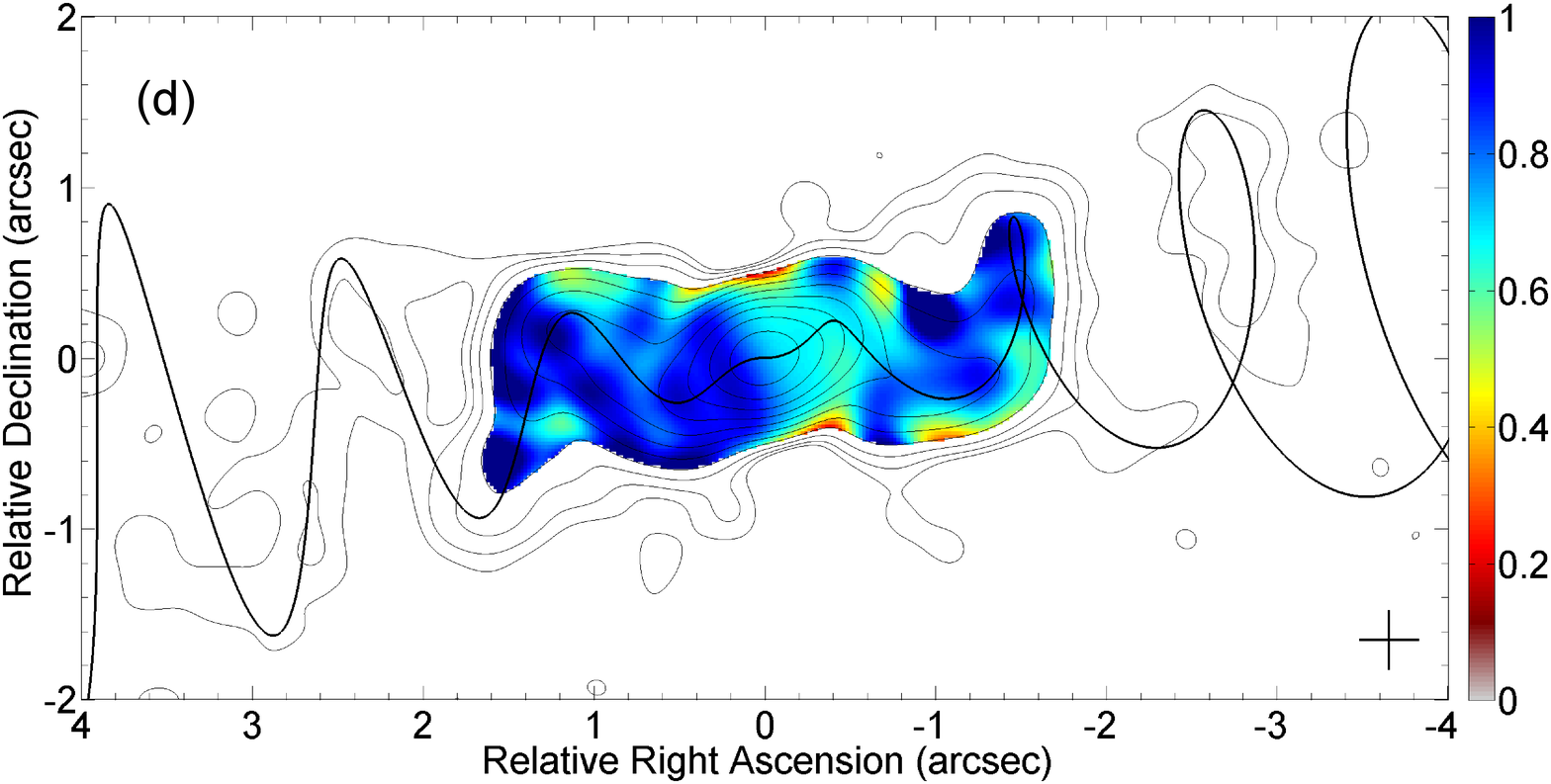}
	\caption[Spectral Index Maps]{Maps of the spectral index $\alpha$ ($\mathrm{S}_\nu \propto \nu^{-\alpha}$). The contours are of C-band total intensity with steps of 2 between levels.  The restoring beam was specified to be circular with full-width half-max $0.350\arcsec$. The kinematic model is displayed over each image. (a) 2007 June 8, (b) 2007 July 1, (c) 2007 July 18, (d) 2007 August 5.}
	\label{fig:spectral_index}
\end{figure}

The resulting maps are shown in Figure \ref{fig:spectral_index}. Each has been clipped to include data only where the estimated spectral index uncertainty is less than 0.1. In each map, the uncertainty in $\alpha$ ranges between 0.05 and 0.1. This estimate includes map noise and the $3\%$ calibration uncertainty. The contours are of total intensity at 5 GHz and increase in steps of $2$. The kinematic model is displayed over each image. The spectral index map for the August 24th epoch has not been included.  Excessive errors in the images at this epoch, particularly around the core (see the large diagonal streaks in Figures \ref{fig:cuni} and \ref{fig:xuni}) prevented us from measuring the spectral index accurately from these images.

In addition to matching the image resolutions, we must also consider the fact that the UV coverage between the 5 and 8.5 GHz observations is not the same, particularly at short spacings.  In any interferometric observation there is a 'hole' in the center of the UV plane that corresponds to the minimum distance between antenna pairs.  For the observations presented here the minimum spacing is $\sim 3$ k$\lambda$ at 5 GHz and $\sim 6$ k$\lambda$ at 8.5 GHz. 

We tested whether missing short spacing flux at 8.5 GHz makes a significant change in the spectral index images in two ways. First, we simply removed visibilities at spacings shorter than 6 k$\lambda$ from the 5 GHz data to match the minimum spacings at 8.5 GHz. Second, we used the AIPS task UVSUB to sample a CLEANed C-band image with the UV coverage of the X-band observation at the same epoch. Each modified UV data set was then re-imaged and compared to the original C-band image to check for a decrease in intensity due to the missing short spacings.  In neither case were the images or resulting spectral index maps significantly changed.

We find that, within measurement uncertainties, the spectral index is uniform across the source; there is no evidence of significant spatial structure or evolution in time. The average spectral index is found to be $0.68\pm0.12$, $0.73\pm0.10$, $0.77\pm0.13$, and $0.79\pm0.15$ for June 8th, July 1st, July 18th, and August 5th respectively. From these measurements we conclude that the average spectral index is $0.74\pm0.06$, consistent with the value of $0.6-0.8$ measured by \citet{stirling04}, only slightly steeper than the 0.6 measured by \citet{1980ApJ...241L..77S}, and consistent with the value of $0.7$ used in Paper~II.

We find no evidence for a flattening of the spectrum at the core.  On VLBI scales, \citet{paragi99} found an inverted spectrum at the base of the jets, which are separated by an AU-scale gap that scales inversely with frequency as predicted by the model of \citet{blandford79}. The jets observed in their images are unresolved in our images; the entire VLBI scale jet is contained within the VLA-scale core.  Since only the base of the miliarcsecond-scale jets has an inverted spectrum and the spectrum quickly steepens, we should expect the core in our images to have a similar spectral index to the rest of the jet.


There is also no evidence of steepening due to radiative losses on this scale. The half life of a synchrotron radiating electron is \citep[][eqn 3.57]{deyoung02}

\begin{equation}
	\tau_{s}=(470\mbox{ yr})\left(\frac{B}{10\mbox{ mG}}\right)^{-3/2}\left(\frac{\nu}{5\mbox{ GHz}}\right)^{-1/2}.
\end{equation}
The fiducial field strength is taken from \citet{bell_thesis} who measured an equipartition field strength in the jets of SS\,433 of $10\mbox{ mG}$ at a distance of $\sim1\arcsec$ from the core. Therefore the lifetime of the radiating electrons is much longer than the age of the jet material at these scales (less than one year). We would have to observe $\sim1100$ precessional wavelengths along the jet in order to measure any steepening at our observation frequencies. This is comparable to the time it takes for the jet material to reach the shell of W50, the supernova remnant in which SS\,433 is contained \citep{elston87, dubner98}.

\section{The Intrinsic Brightness Profiles}\label{sec:intrinsic_brightness}


The analysis of radio maps of even mildly relativistic jets is complicated by Doppler beaming. For SS\,433, the helical geometry also requires that consideration be given to projection effects. Fortunately the geometry of SS\,433 is well known and we are able to correct for these effects in order to estimate the intrinsic properties of the jets. In Paper~II we developed methods for calculating the intrinsic brightness and applied them to a very deep 5 GHz VLA image of SS\,433 made from an observation in 2003 July. With the data presented here we can repeat the same calculations on a series of deep images thus allowing us to obtain corresponding results at several observational epochs. Additionally we can study the dependence of the intrinsic brightness not only on the age of the material when it emits the radiation that we observe, but also on the birth epoch of the material.

Before describing the corrections we should briefly explain what we actually want to measure. What does the term ``intrinsic brightness'' mean specifically? If we describe the jets as if they were composed of many closely spaced individual components, we would like to know the power emitted by each component as measured by a co-moving observer. If we describe the jets as continuous, we would like to know the emitted power per unit length along the jet helix.  

As a result of the (mildly) relativistic jet velocity, we must consider the effects of Doppler beaming.  A quantitative measure of this effect requires knowledge of the velocity vector at each location along the jet helix.  Fortunately the necessary velocities and angles are easily obtained from the kinematic model.  Additionally, because of the helical geometry of the jet and the distortion of the observed helix due to differences in light travel time between different parts of the jet, the amount of material that contributes flux to the observing beam will differ from location to location. In order to determine the contribution per unit length (or per component) we must also consider the changing projected density on the sky. Again using the model, we can calculate the density and normalize the measured brightness by this value.

We will refer to four different time coordinates that must be defined explicitly. The birth epoch, i.e. the date at which a part of the jet, as seen by the observer, was emitted from the core, will be denoted as $t_{b}$ (we will always list the birth epoch using a Truncated Julian Date, TJD $=JD-2454000$). The time $t$ denotes the observation date for a distant observer (our radio telescope). We also make use of the ``age'' of a piece of the jet, that is, how long it has been since the piece in question emerged from the core. The observed age of a piece of the jet is defined as $t_{age}=t-t_{b}$ (assuming a constant velocity). Because the light travel time varies along the jet helix, the age of a piece of the jet when it emits the observed photons is generally not equal to the observed age. The relationship between the observed age of a piece of the jet ($t_{age}$) and the age of the same jet material when it emitted the photons we receive ($\tau$) is
\begin{equation}
	\tau=\frac{t_{age}}{1-\beta\cos\theta}=\frac{t_{age}}{1-v_{x}/c}
\end{equation}
where $\theta$ is the angle between the local velocity vector and the line of sight. This is the same $\tau$ used in Paper~II and called ``age at time of emission of photons''.

If the model information is stored in a list containing $N$ model jet components equally spaced in $t_{b}$, the total correction factor, $C_{i,n}$, at the position of component $i$, including both projection and relativistic effects, is 
\begin{equation}
	C_{i,n}=\sum_{j=1}^{N}D_{j}^{n+\alpha}e^{-(\Delta R_{ij})^{2}/(2\sigma^{2})}.
\end{equation}
Here $D=[\gamma(1-\beta\cos\theta)]^{-1}$ is the Doppler factor, $\sigma$ is the standard deviation of the Gaussian CLEAN beam ($FWHM=2\sqrt{2\ln2}\sigma$), and $\Delta R_{ij}$ is the angular separation between components $i$ and $j$. We normalize this correction to unity at the core. Here $n$ is a parameter that is $2$ for a continuous jet and $3$ if the jet is composed of individual, resolved components. In order to recover the ``intrinsic brightness profile'' of the jet, we simply take a reading from the image at the location of each model point and divide by this (dimensionless) correction factor. 

This correction assumes that each part of the jet within a beam area has the same intrinsic luminosity. This is approximately true for most of the jet, but fails at the crossing points of the loops on the west. Here, material of very different emission ages contributes flux to the same beam. As a result, the calculated intrinsic brightness at the front of the loop (corresponding to the younger jet material within the beam) will be slightly lower than it should be, while that of the older components at the rear of the loop will be higher than it should be. Without information about how the brightness evolves in $\tau$ and varies with $t_b$ it is impossible to attribute the appropriate amount of brightness to each part of the jet.

\begin{figure}
	\centering
	\plottwo{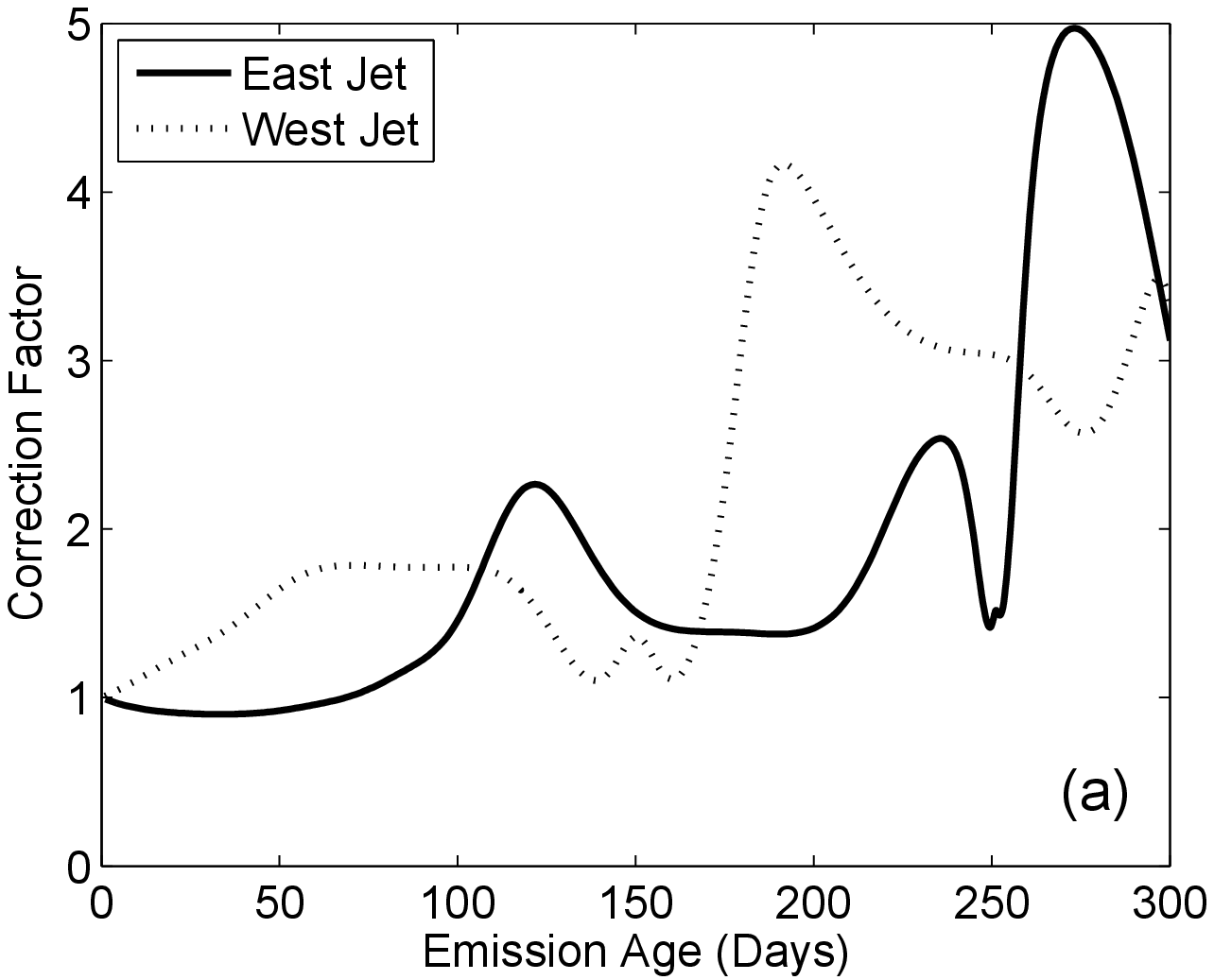}{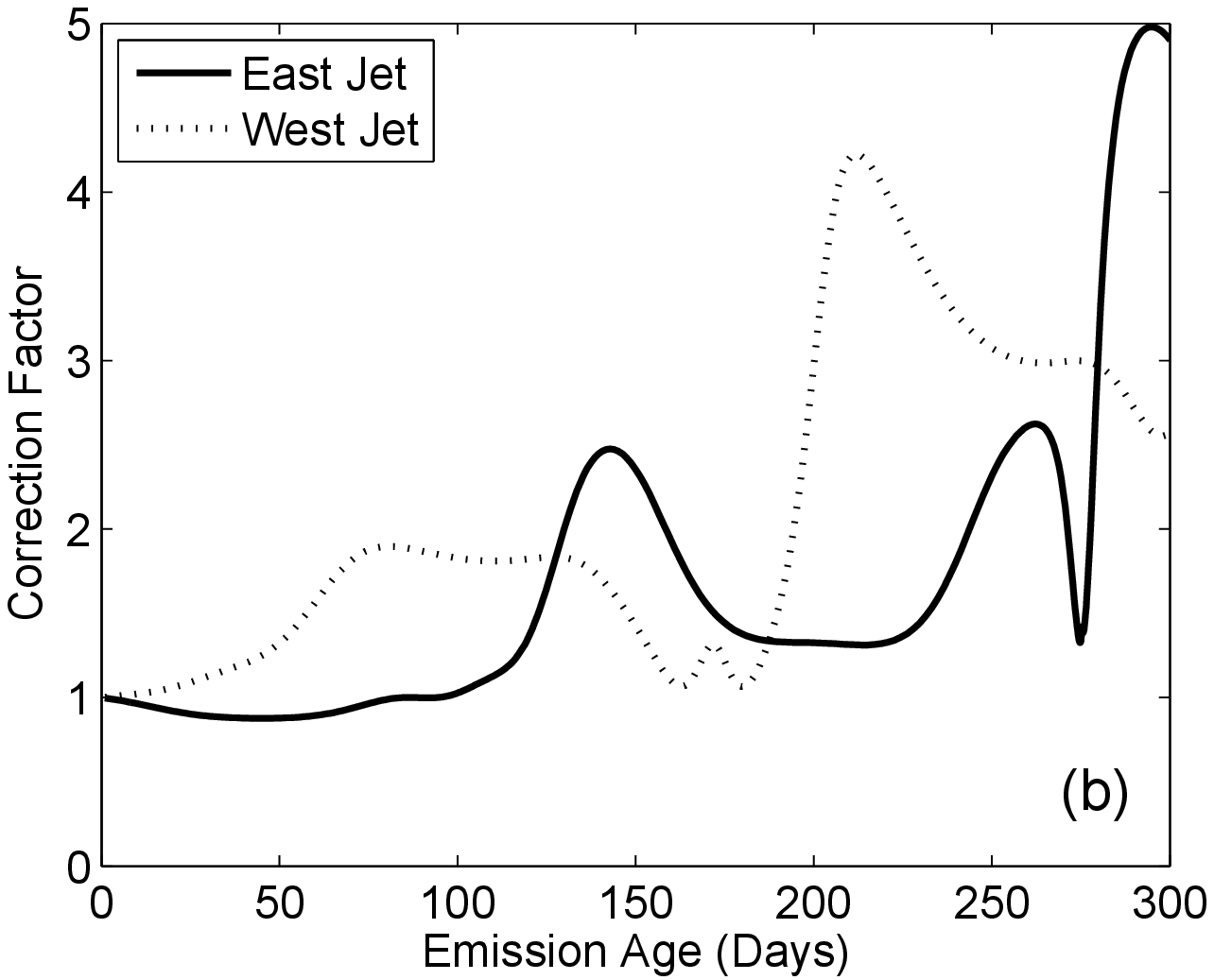}
	\plottwo{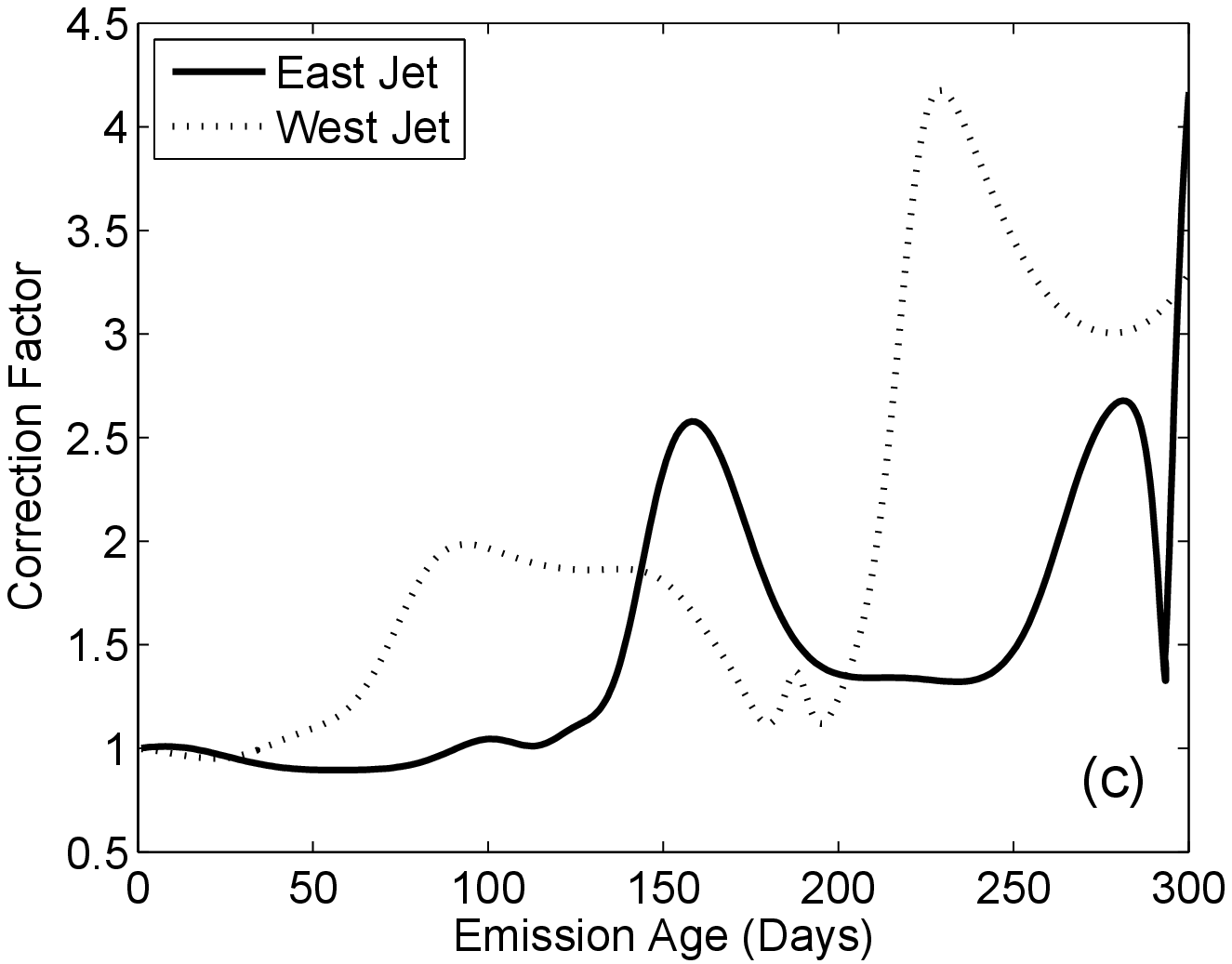}{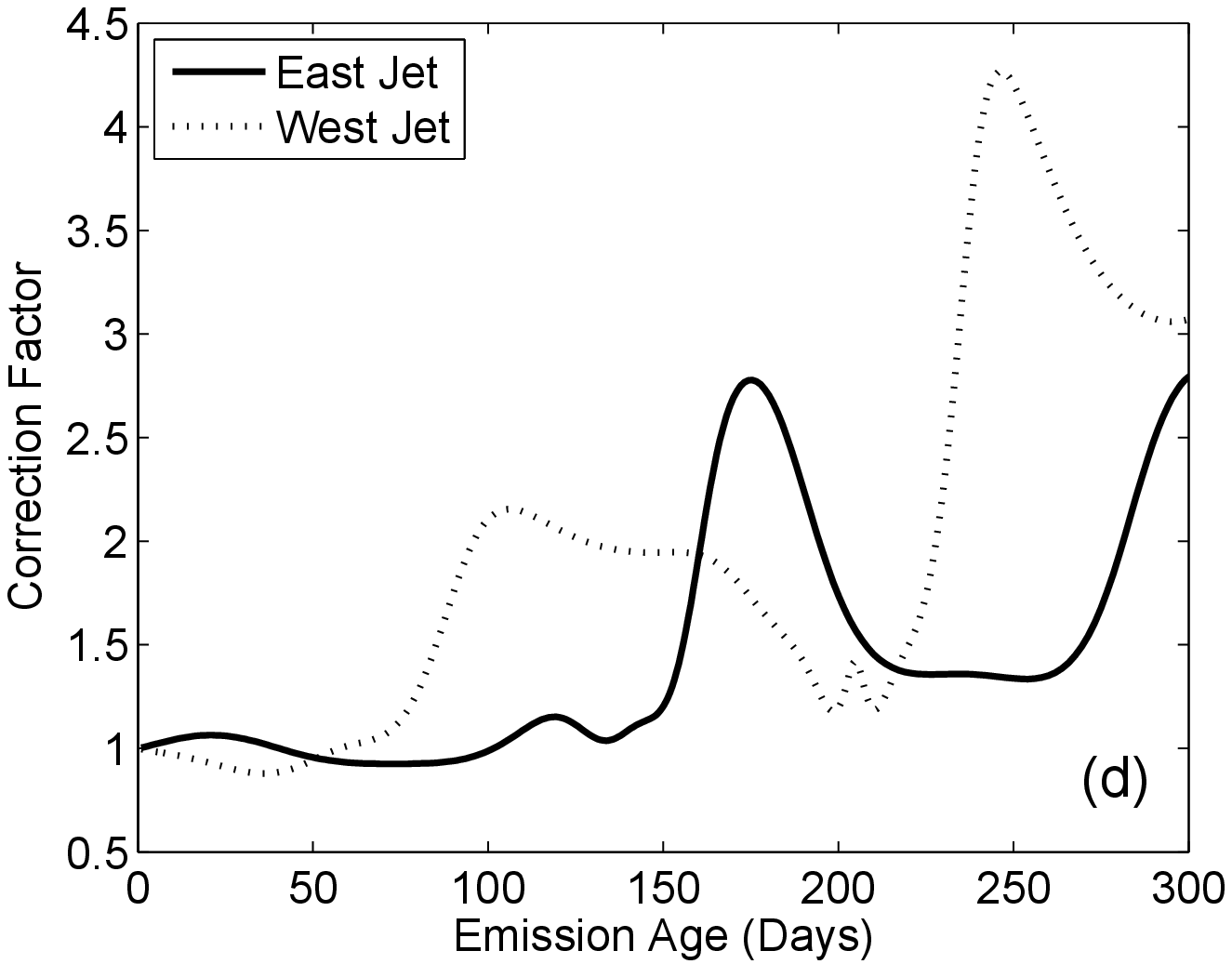}
	\epsscale{0.5}\plotone{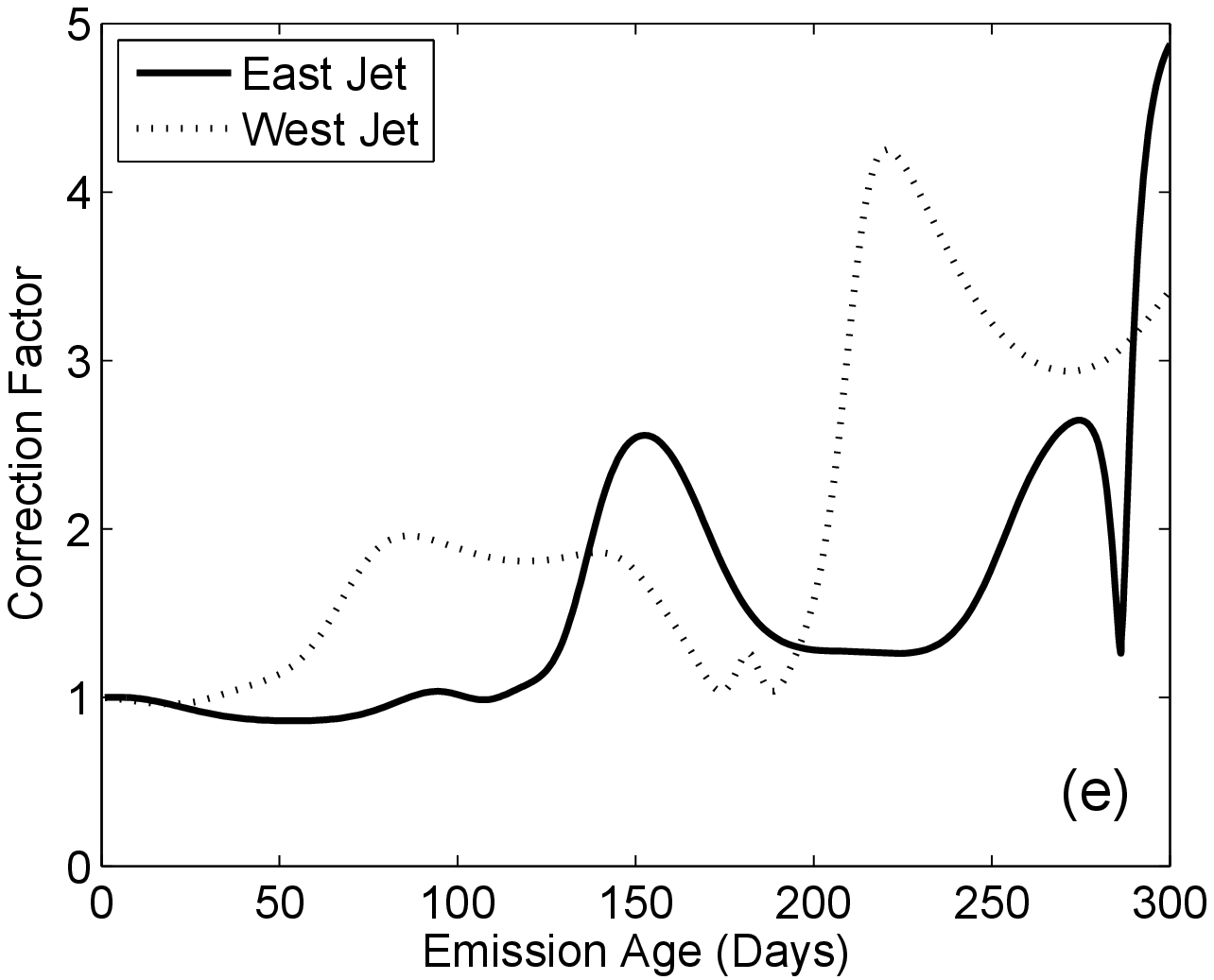}\epsscale{1.0}
	\caption[Correction Factors]{The multiplicative correction factors. The intrinsic brightness is the product of this correction factor and the observed total intensities. (a) 2007 June 8, (b) 2007 July 1, (c) 2007 July 18, (d) 2007 August 5, (e) 2007 August 24.}
	\label{fig:corr_factors}
\end{figure}

The question of the choice of $n$ was addressed in some detail in Paper~II where we compared profiles from a 2003 VLA image at 5 GHz that had been normalized with each value. We determined that $n=2$ is the appropriate value because the jet profiles appear more symmetric and, most convincingly, the observed west jet profile was very accurately reconstructed from the intrinsic east jet profile. Similarly, here with the 2007 data we find that the results are more consistent with a value of $n=2$ and will use it in the following analysis. Figure \ref{fig:corr_factors} shows $1/C_{i,2}$, the factor by which each sample is multiplied, as a function of $\tau$ for each epoch.  

\begin{figure}
	\centering
	\plottwo{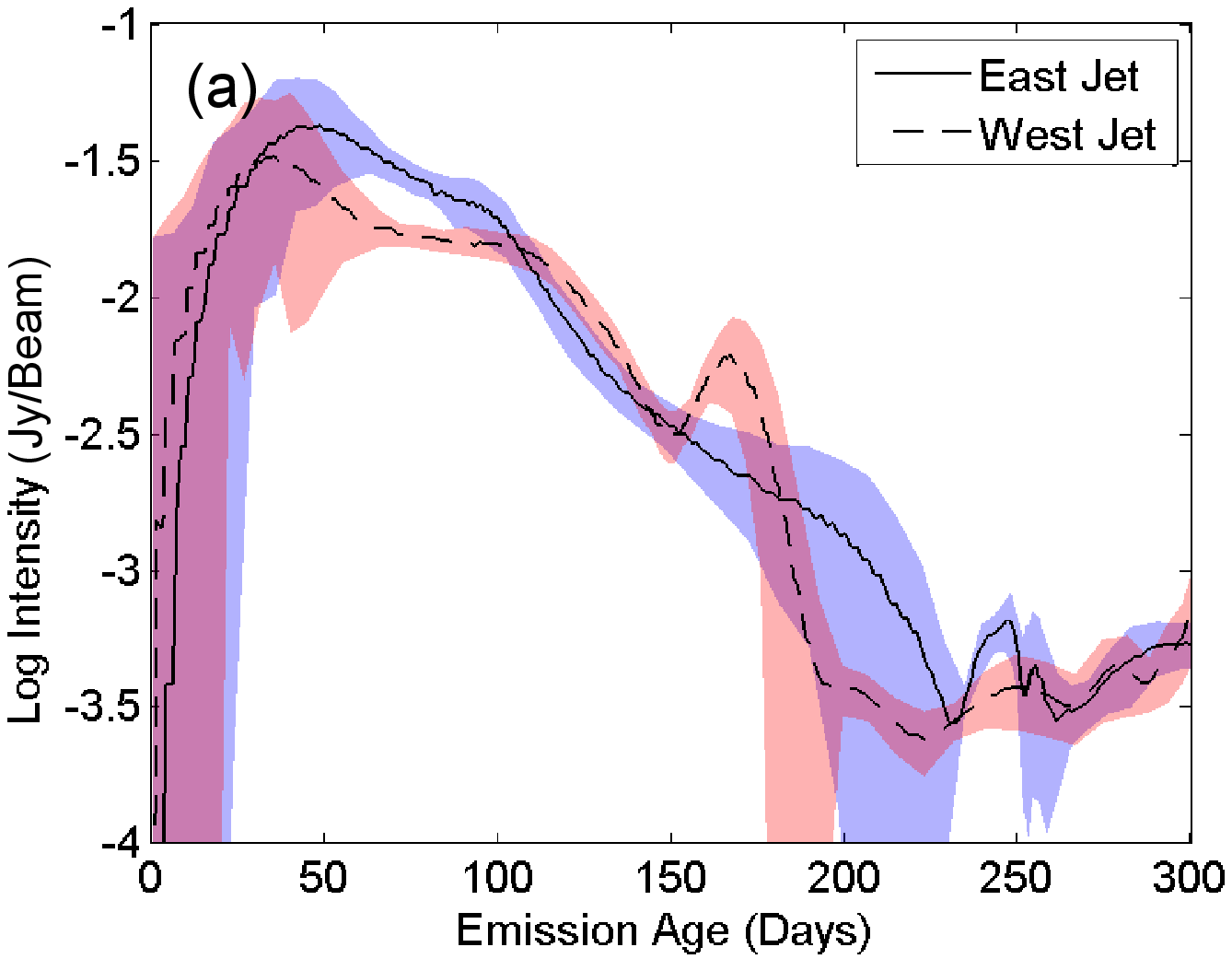}{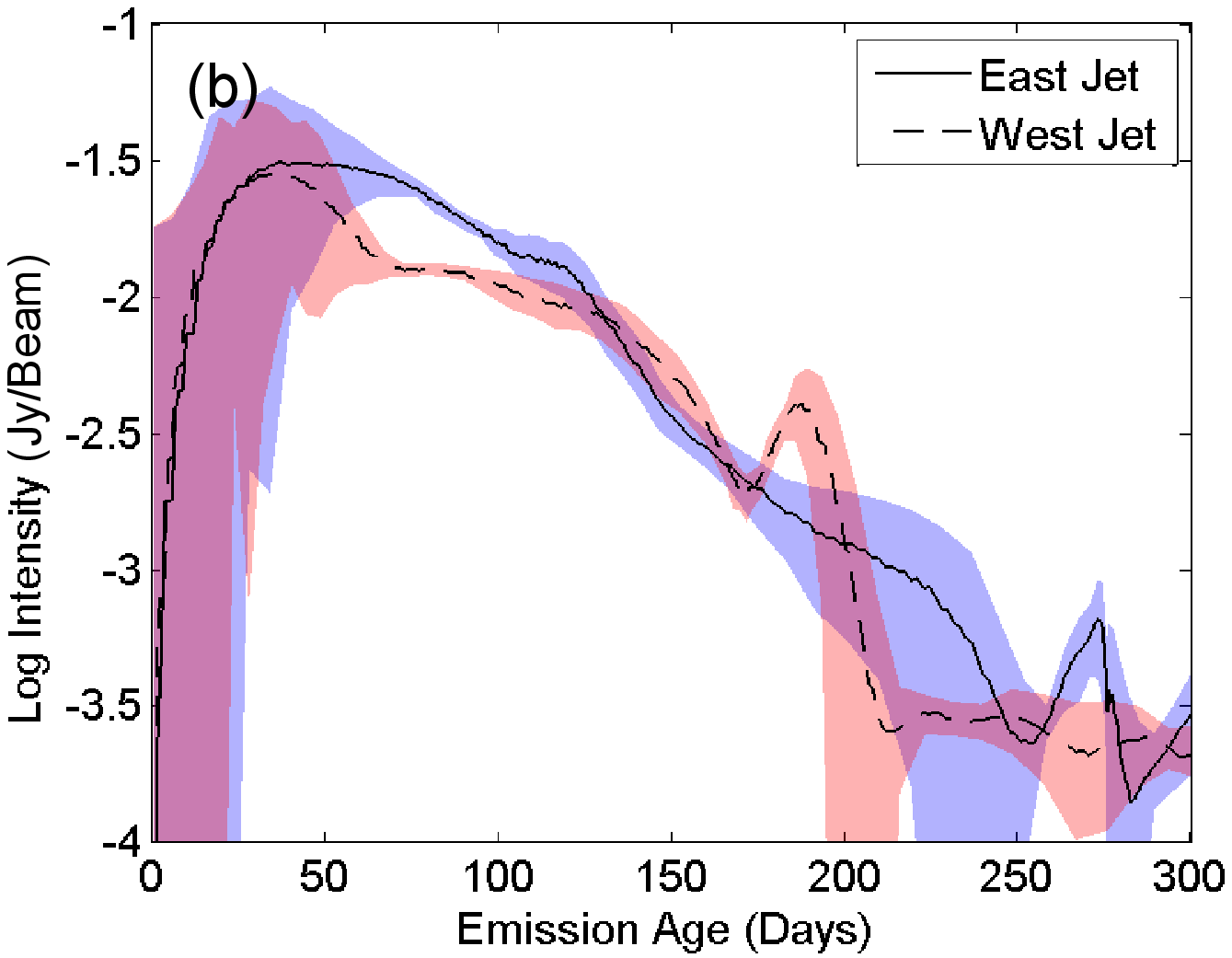}
	\plottwo{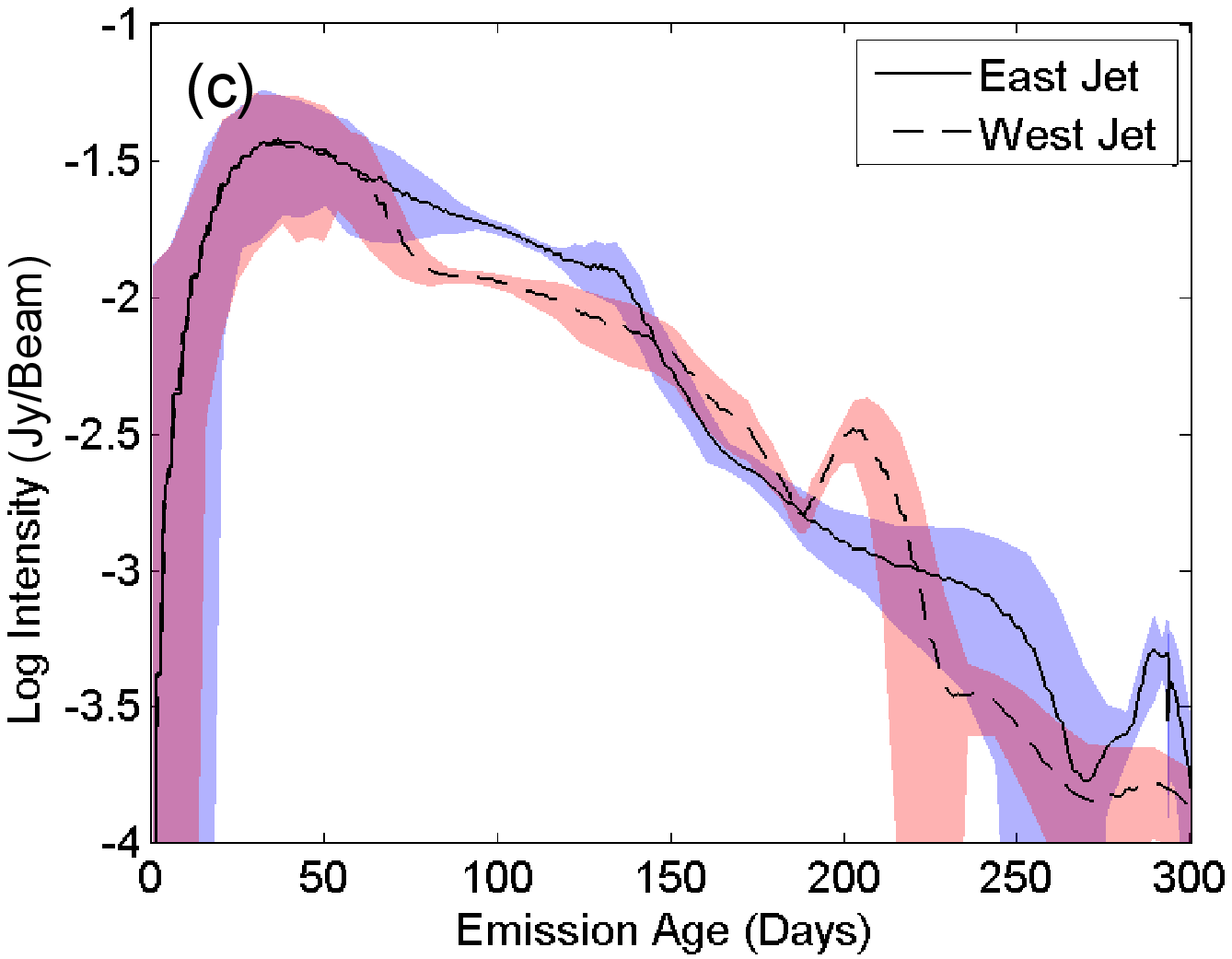}{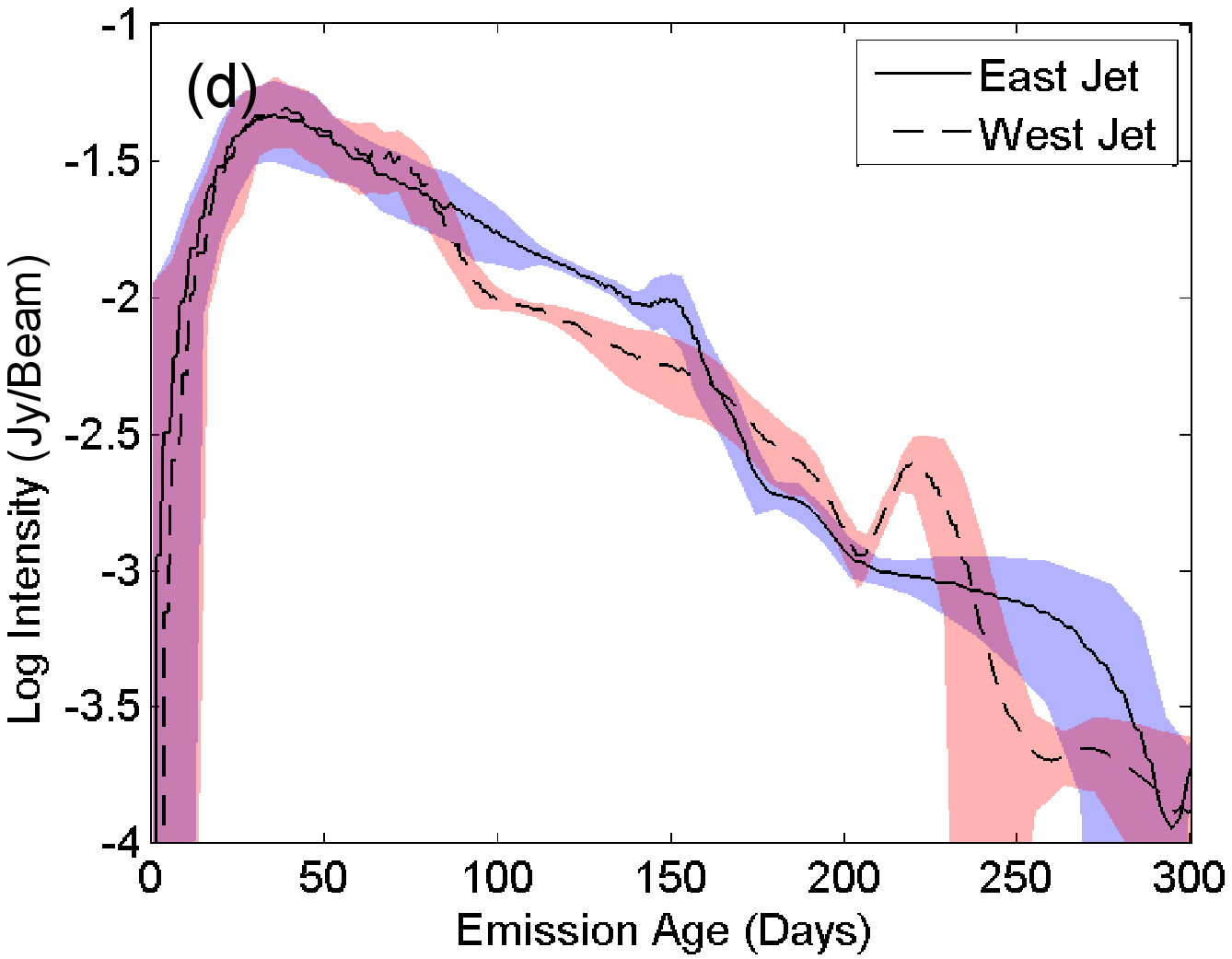}
	\epsscale{0.5}\plotone{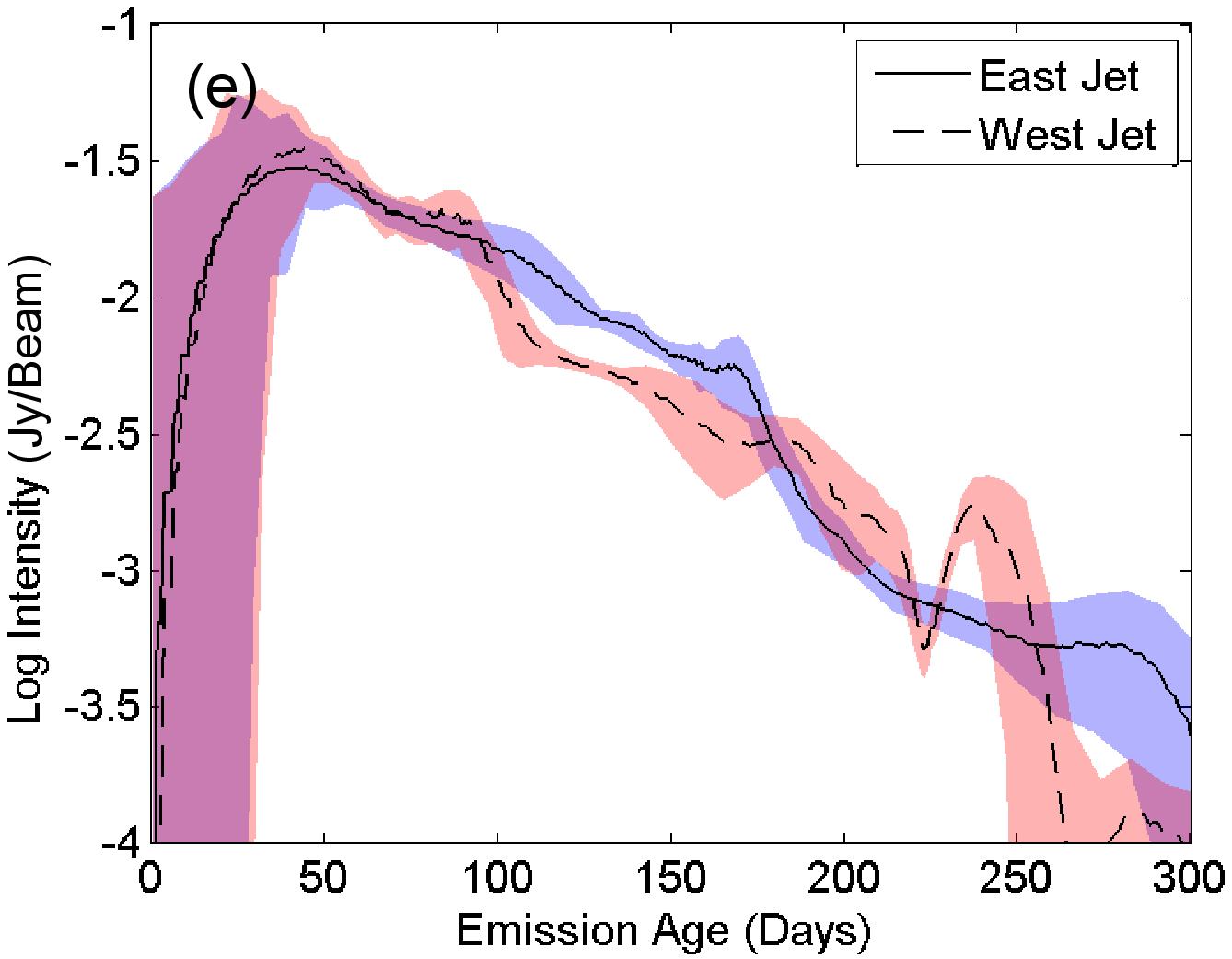}\epsscale{1.0}
	\caption[Measured Brightness Profiles]{Observed total intensity profiles from the C-Band observations as a function of the emission age. The colored regions indicate the estimated uncertainties; blue for the east jet and red for the west.  The core was subtracted from each image prior to measuring the profiles. The bump in the west jet profiles that moves from $\sim$150 to 250 days is where the projected helix crosses itself (see text). (a) 2007 June 8, (b) 2007 July 1, (c) 2007 July 18, (d) 2007 August 5, (e) 2007 August 24.}
	\label{fig:profiles_raw}
\end{figure}

Figure \ref{fig:profiles_raw} shows the measured intensity profiles from each of the five 5 GHz uniformly weighted images. The 8.5 GHz profiles are similar and not shown. These have been plotted against emission age $\tau$ rather than the observed age (which includes light travel time) because it is $\tau$ that defines the evolution of the system. A Gaussian component having the same dimensions as the restoring beam has been subtracted from the images at the core prior to measuring the jet brightness which is why the profiles fall precipitously to zero as $\tau$ approaches zero. 

The blue and red shaded areas on the plots represent the uncertainties of the measurement of flux density in the east and west jets, respectively. To estimate these uncertainties we consider the noise level of the image, a three percent overall flux calibration uncertainty (which is a conservative estimate), and the uncertainty in model position. To calculate this last source of uncertainty we measure the flux density of the map using three models, one created using the mean speed of $\beta=0.2647$ and the other two with speeds differing by $\pm 10\%$ from the mean. \citet{blundell04} report a 10\% fluctuation in velocity, so this should give a reasonable measurement at the extremes of the possible model positions. We then measure the average deviation in the measured values from each of these models and take it to be the magnitude of uncertainty in the flux density due to uncertainty in the position of the jet components. This source of uncertainty dominates over the first two; the RMS difference between the profiles measured using different models is $\sim0.5$ mJy which is roughly an order of magnitude greater than the map noise and calibration uncertainty.

When studying these plots it is helpful to note the values of $\tau$ for which the local velocity vector is in the plane of the sky. Table \ref{tab:reference_times} lists these times as well as the times when the jet velocity vector makes maximum and minimum angles to the line of sight. Careful inspection of the observed jet profiles (Figure \ref{fig:profiles_raw}) shows that at the sky crossing times the brightness of the east and west jets is very similar, as expected since the Doppler factors and projection at those times are the same. At other times there appear to be asymmetries in the measured brightness, and the east and west profiles in general have rather different shapes. For instance, there is a large difference between the east and west intensity values between 50 and 100 days in the early observations. The asymmetries and features in the curve, such as the sharp break at 100 days in the June 8th profile, appear in the profiles of each epoch at later times as the summer goes on. The aforementioned difference, for instance, moves outward to 100 to 170 days by August 24th.

\begin{figure}
	\centering
	\plottwo{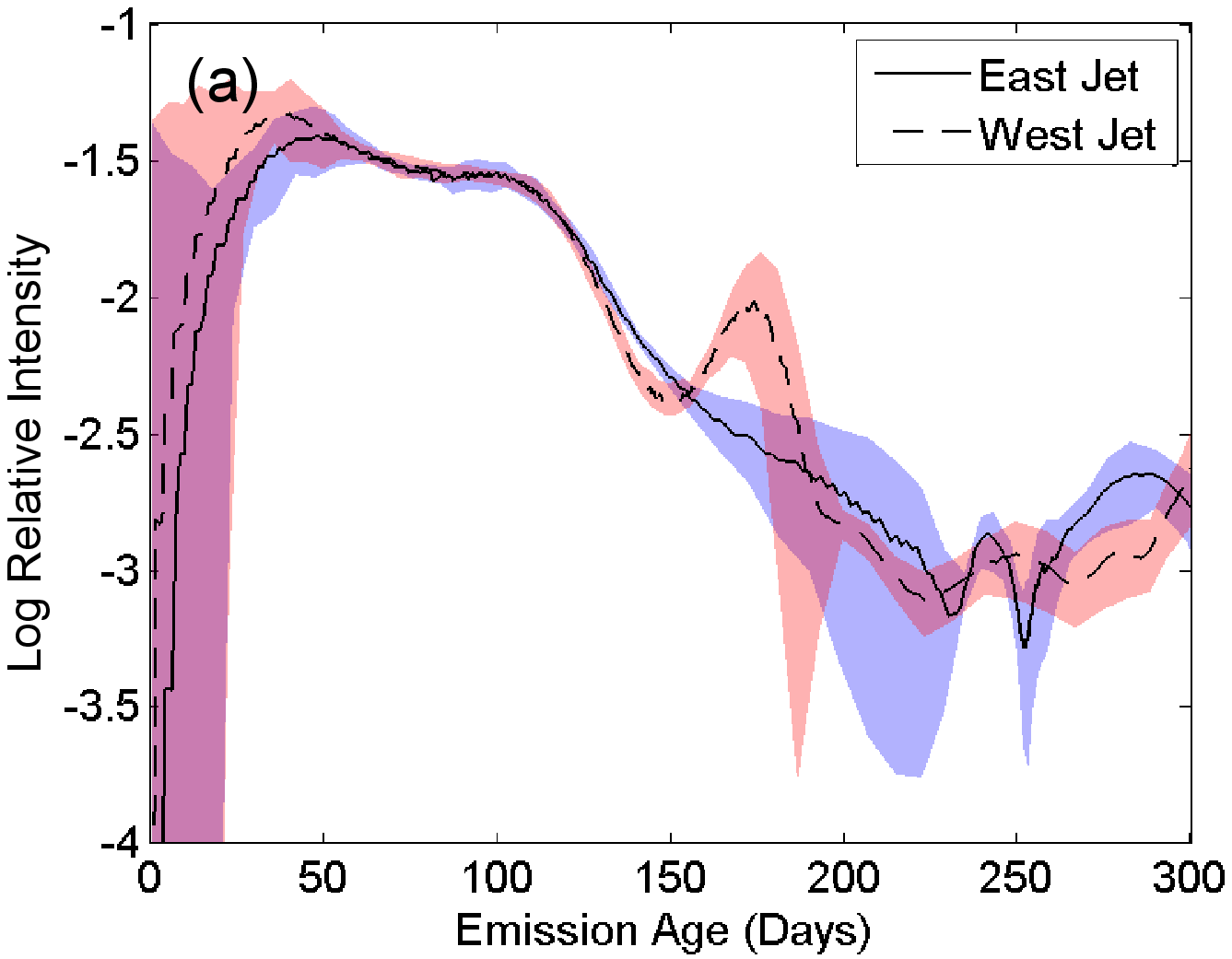}{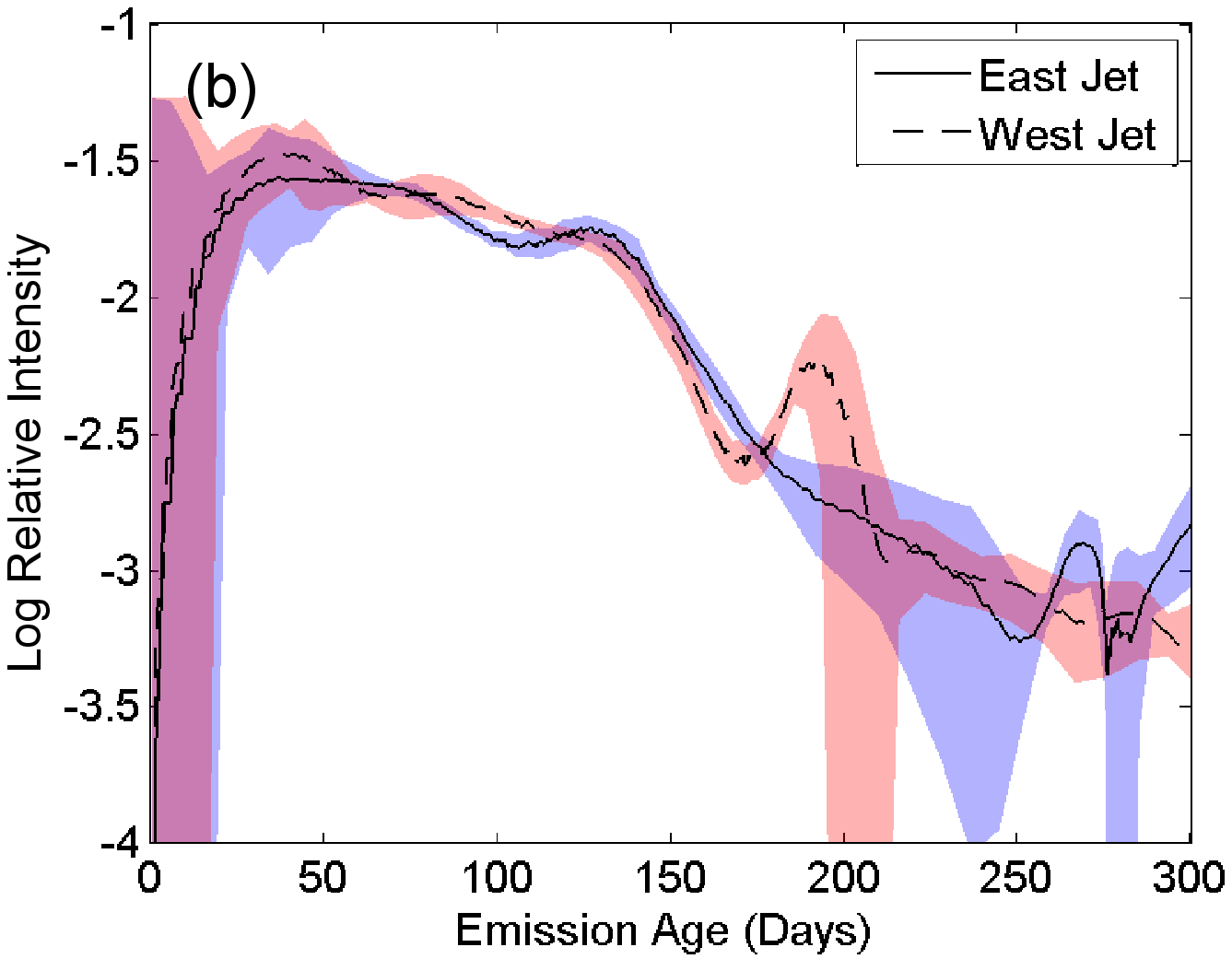}
	\plottwo{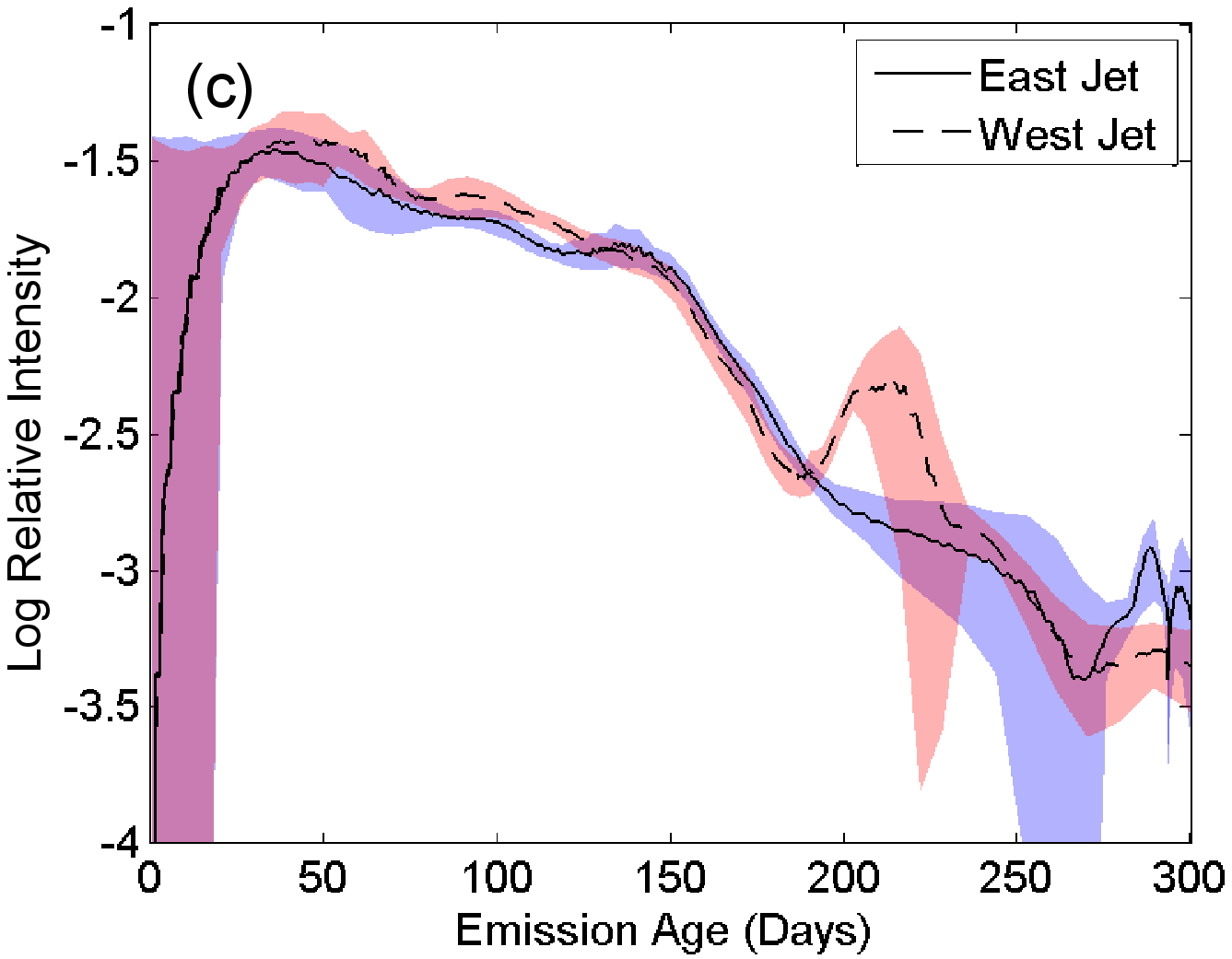}{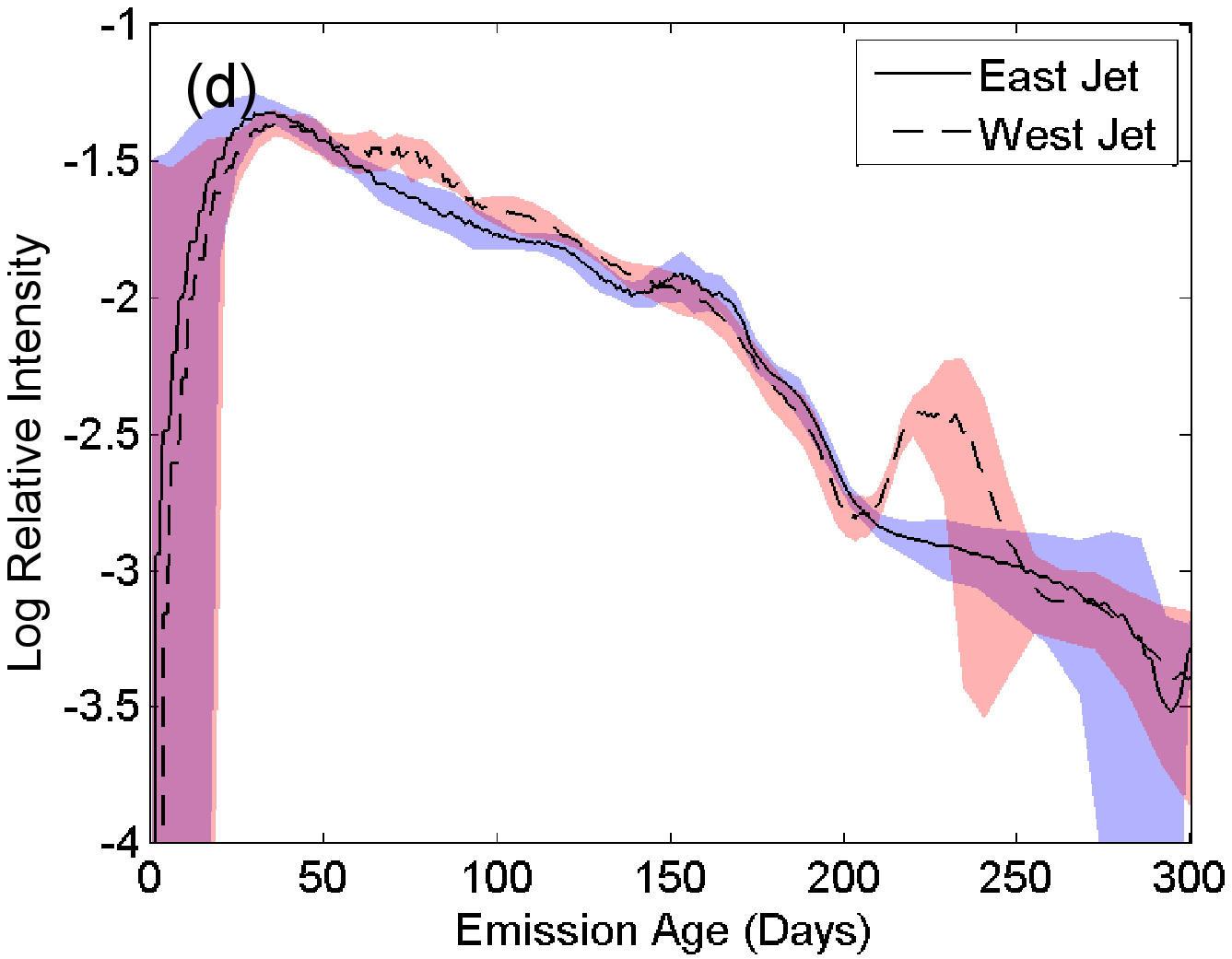}
	\epsscale{0.5}\plotone{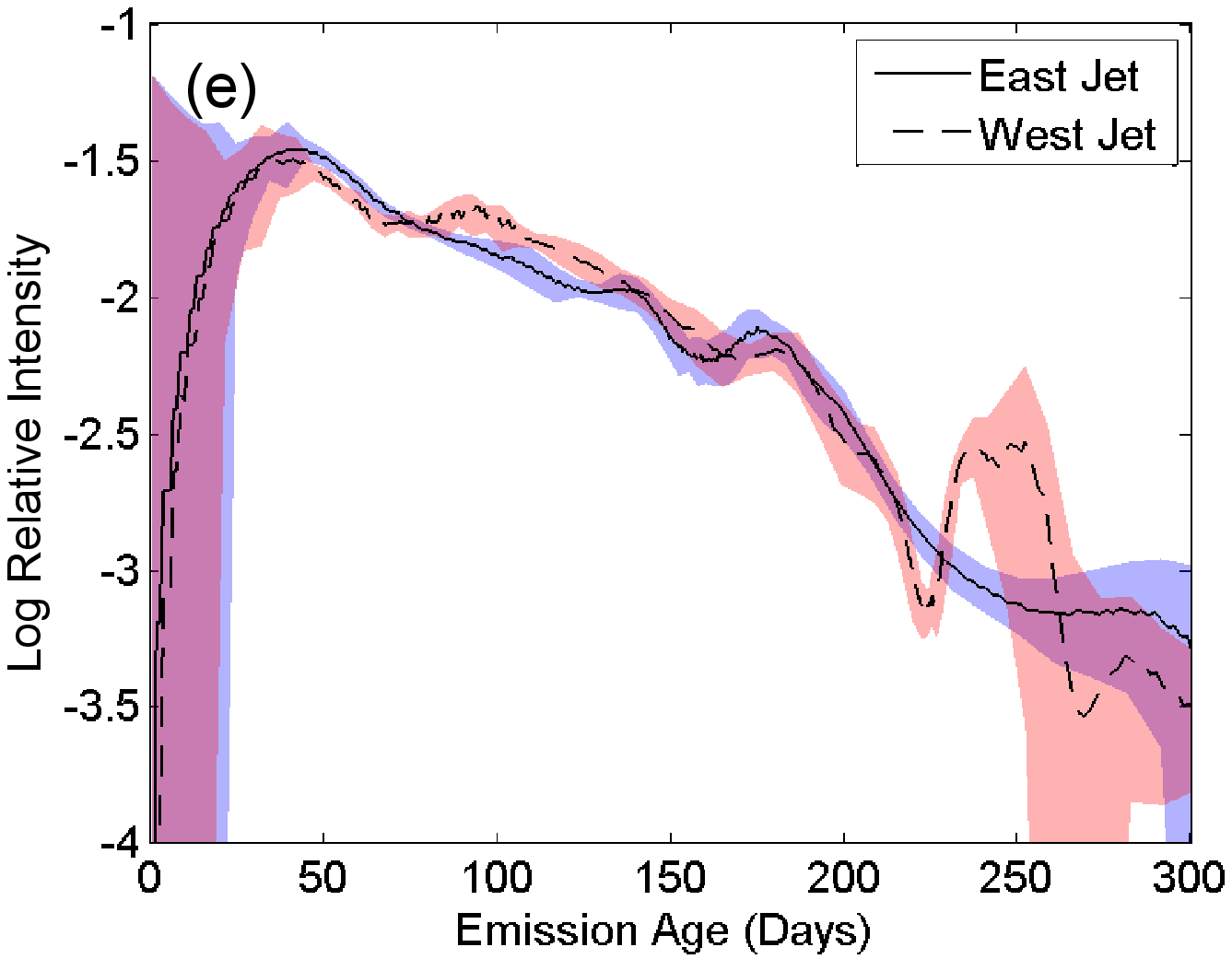}\epsscale{1.0}
	\caption[Intrinsic Brightness Profiles]{The intrinsic brightness profles of the C-Band observations as a function of the emission age. The colored regions indicate the estimated uncertainties; blue for the east jet and red for the west. The core was subtracted from each image prior to measuring the profiles. The bump in the west jet profiles that moves from $\sim$150 to 250 days is where the projected helix crosses itself (see text). (a) 2007 June 8, (b) 2007 July 1, (c) 2007 July 18, (d) 2007 August 5, (e) 2007 August 24.}
	\label{fig:profiles_pb2}
\end{figure}

Figure \ref{fig:profiles_pb2} shows the intrinsic brightness profiles. The corrected profiles appear much more symmetric, with the only major difference being a small dip followed by a large bump in the west jet profile. This is seen at $\tau\sim150-170$ days in the June 8th profile and appears at later times in subsequent images. As described above, this apparent mismatch results from components of very different ages contributing flux to the map at the loop crossing point in the west jet (see, for instance, the region at relative right ascension $-1.2\arcsec$, declination $0.5\arcsec$ on the July 1st image in Figure \ref{fig:cuni}). The minimum of the small dip coincides with the front of the jet cone while the maximum corresponds to the value measured at the location of the ``older'' model point at the rear of the jet cone.

There are other, smaller differences between the east and west jet intensities in the corrected profiles. For instance, around $\tau=100$ days in the July 1st profile, the intrinsic brightness measured in the west jet is $20\%$ higher than that measured in the east jet. This difference is slightly outside of the estimated uncertainties. However, for this value of $\tau$ there is a difference in $t_{b}$ between the east and west jet of $\sim10$ days. There are similar differences at the other epochs as well, but they all appear when there is a significant difference in birth epoch between the east and west jet components at a given $\tau$. In each profile, where $\Delta t_{b}=0$ the profiles match quite well. We conclude that the apparent difference between profiles is indicative of intrinsic flux variation with $t_{b}$. We investigate this further in the next sections.

\subsection{Evolution of the Intrinsic Brightness}\label{ssec:evolution}

When examining the intrinsic jet profile of a single epoch as we did in Paper~II it is immediately clear that the curve is not well described by a simple mathematical function. From the 2003 July 11 VLA observation we found three regions, each best described by different power law or exponential functions. In the range $50 \lesssim \tau \lesssim 150$ days, the intrinsic brightness curve was adequately fit by a power law with exponent $-1.8$ or an exponential function with a half-life of $\sim40$ days. The curve then flattened, displaying a roughly constant intrinsic brightness from $150\lesssim \tau \lesssim 300$ days. Beyond $\sim 300$~days the intrinsic brightness fell off more quickly, and was best fit by a power law with an exponent of roughly $-4$ or an exponential function with a half-life of $\sim80$ days. 

\citet{hjellming88} also noted a break in the jet brightness profile, with the rate of decay increasing sharply at $t_{age}=100\mbox{\,\ days}$. Based on their model of a precessing conical jet (discussed below), they attributed this behavior to a change in the transverse jet expansion rate. They argued that the jets initially undergo confined expansion and then, after about 100 days, the jets expand freely. 

\begin{figure}
	\centering
	\plotone{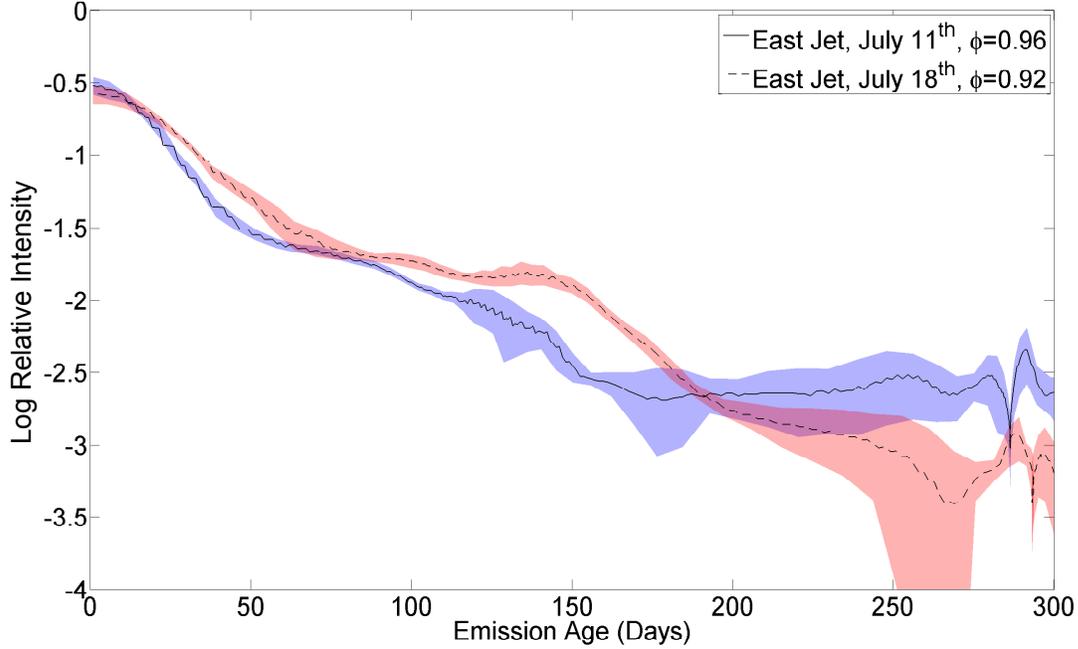}
	\caption[July 11th vs. July 18th]{Comparison between the east jet intrinsic brightness profiles of 2003 July 11 and 2007 July 18.}
	\label{fig:711_v_718}
\end{figure}

Similar breaks are apparent in the profiles of the 2007 data presented here. From our multi-epoch data set we see that these features propagate outward from epoch to epoch. The brightness curves from the 2007 images are quite different from that of the 2003 epoch. Figure \ref{fig:711_v_718} compares the intrinsic east jet profiles of the 2003 July 11 epoch (Paper~II) and the 2007 July 18 epoch. On these dates the source was at a similar precessional phase, 0.92  in 2003 and 0.96 in 2007. 

Observations of the jet profile from a single epoch are insufficient for characterizing the evolution of the jet brightness since, in general, the brightness of any piece of the jet might be expected to depend on two time parameters. These are $t_{b}$, the birth epoch, and $\tau$, the age of the component when it emits the photons that we observe. The dependence on $t_b$ reflects the varying level of power injected into the jets at the core, and the dependence on $\tau$ reflects the evolution of individual pieces of jet. From the jet profile at a single epoch we can measure the brightness of the jet for many different values of $t_b$, but we only measure the brightness for one value of $\tau$ at each value of $t_b$. With the multi-epoch observations presented here, however, we have a measure of the intrinsic brightness at up to five values of $\tau$ for each $t_b$ (ten if we assume east/west symmetry). 

\begin{figure}
	\centering
	\plottwo{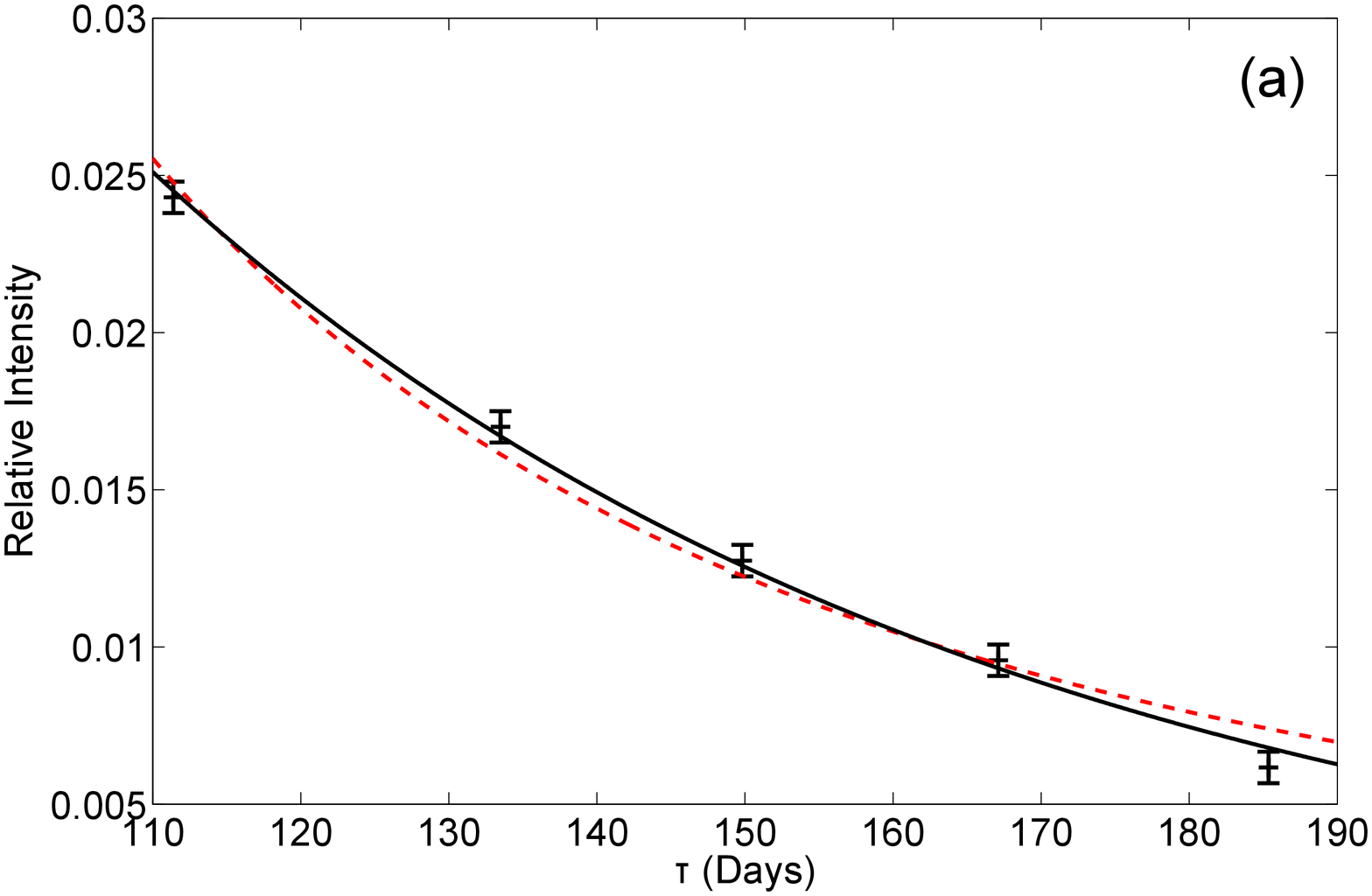}{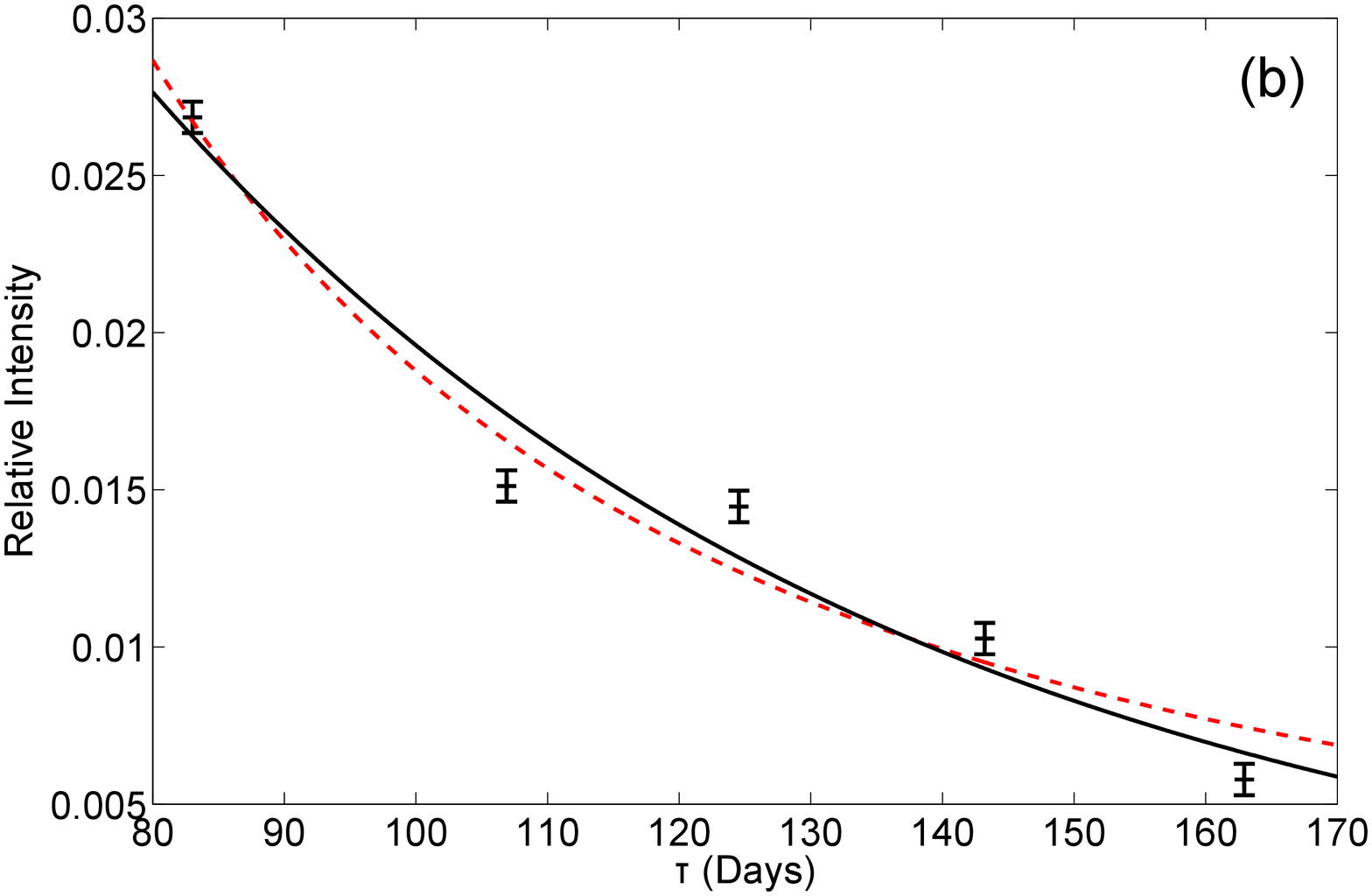}
	\caption[Individual brightness measurements at fixed $t_b$ as a function of $\tau$]{The evolution of the C-band intrinsic brightness for two pieces of the eastern jet emitted at different birth epochs ($t_b$). The uncertainty is estimated using the procedure employed for Figures \ref{fig:profiles_raw} and \ref{fig:profiles_pb2}. The best power law and exponential fits are shown as dashed red and solid black lines, respectively. (a) Fits for $t_b = 143$: $\tau'=58\pm6$ days and $s=2.4\pm0.5$, (b) fits for $t_b = 179$: $\tau'=58\pm10$ days and $s=1.9\pm0.8$.}
	\label{fig:fits-segments-v-component}
\end{figure}

We will assume that the functional form for the brightness at each location along the jet can be factored into parts that each depend only on one of these times,
\begin{equation}
	I(t_{b},\tau) = C(t_b)f(\tau). 
\label{eq:I_is_factorable}
\end{equation}
Under this assumption we can measure $f(\tau)$ if we simply measure the intrinsic brightness at a point with a given $t_{b}$ at each epoch. In Figure \ref{fig:fits-segments-v-component} we show the evolution of the brightness, measured from the east jet, for two different values of $t_b$.  The brightness values in Figure \ref{fig:fits-segments-v-component}a are measured at $t_b=143$ days, which corresponds to $\tau=111$ days and the image location $(0.9, -0.2)''$ on June 8th.  In Figure \ref{fig:fits-segments-v-component}b the brightnesses are measured at $t_b=179$ days, which corresponds to $\tau=83$ days and the image location $(0.5, -0.1)''$ on June 8th. We have chosen points that are well separated on the image and whose intensities have a relatively low uncertainty.  Power law ($\tau^{-s}$) and exponential ($e^{-\tau/\tau'}$) fits are shown as dashed red lines and solid black lines, respectively. For the values measured when $t_b=143$ days we find $s=2.4\pm0.5$ and $\tau'=58\pm6$ days.  For $t_b=179$ we find $s=1.9\pm0.8$ and $\tau'=58\pm10$ days. Both models appear to describe the data equally well. The results are consistent with a constant fitting parameter for either model, although the prediction bounds are quite large due to the limited number of data points used in the fit.

We can improve our description of the evolutionary behavior if we repeat the measurement above for all values of $t_b$. We assume, as suggested by Figure \ref{fig:fits-segments-v-component}, that $f(\tau)$ can be represented by either a power law or exponential function. One of the intensity measurements at a given $t_b$ will be denoted as $I_{0}(t_{b})$ and the corresponding emission age as $\tau_{0}$. Then it follows from equation \ref{eq:I_is_factorable} that the intensity and emission age at other times are related to these by either
\begin{equation} 
	 \frac{I}{I_{0}} = \left(\frac{\tau}{\tau_{0}}\right)^{-s}
	 \label{eq:i_i0_tau_tau0_pow}
\end{equation}
if $f(\tau)$ is a power law, or 
\begin{equation} 
	 \frac{I}{I_{0}} = e^{-(\tau-\tau_{0})/\tau'}
	 \label{eq:i_i0_tau_tau0_exp}
\end{equation} 
if $f(\tau)$ is an exponential.  These ratios have the benefit of being independent of $t_b$ and therefore free from effects of variation in the core. With these ratios we can include all measured intensity values in order to better constrain our best fit parameters, $s$ and $\tau'$, and possibly distinguish between the two models.  

\begin{figure}
	\centering
	\plotone{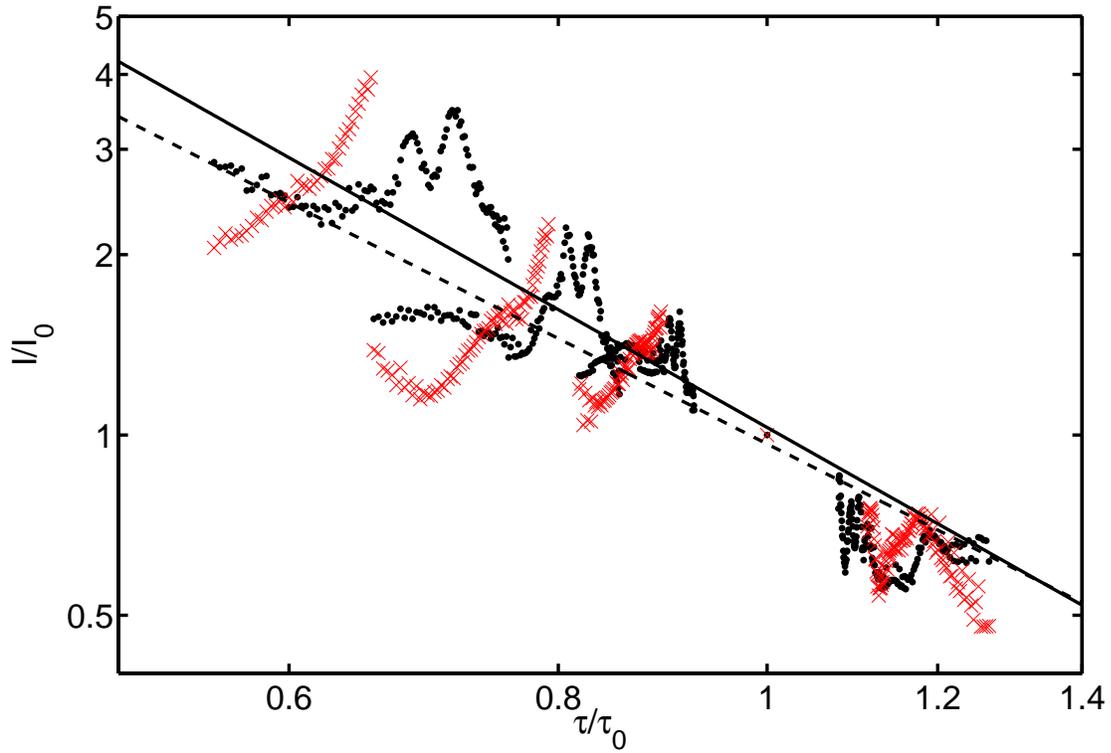}
	\caption[Brightness v. $\tau$ relation assuming power law]{The normalized intrinsic brightness, $I(t_b,\tau)/I_0(t_b,\tau_0)$, as a function of $\tau/\tau_0$.  The trend will be linear if the brightness is a power law in $\tau$.  Values measured from the east and west jet are shown as black dots and red x-marks, respectively.  The best power law fit is shown as the solid line for the east jet and the dashed line for the west jet. For the east jet $s=2.0\pm0.1$, and for the west jet $s=1.6\pm0.1$.}
	\label{fig:ivtau_pow}
\end{figure}

Figure \ref{fig:ivtau_pow} shows the ratio $I(t_{b},\tau)/I_{0}(t_{b},\tau_{0})$ as a function of $\tau/\tau_{0}$. This is plotted on a log-log scale so a power law will appear linear.  This includes intrinsic brightness values from both the east (shown in black) and west (shown in red) profiles. Measurements from the west jet in the loop crossing region, and from regions where the core accounts for more than $1\%$ of the total flux, have been discarded.  The best fit of equation \ref{eq:i_i0_tau_tau0_pow} to the east jet data, shown as a solid line in the figure, yields an exponent of $s=2.0\pm0.1$.  The best fit to the west jet data, shown as a dashed line in the figure, yields an exponent of $s=1.8\pm0.1$.  The R$^2$ (coefficient of determination) values for the fits are 0.76 and 0.70 for east and west, respectively.   

\begin{figure}
	\centering
	\plotone{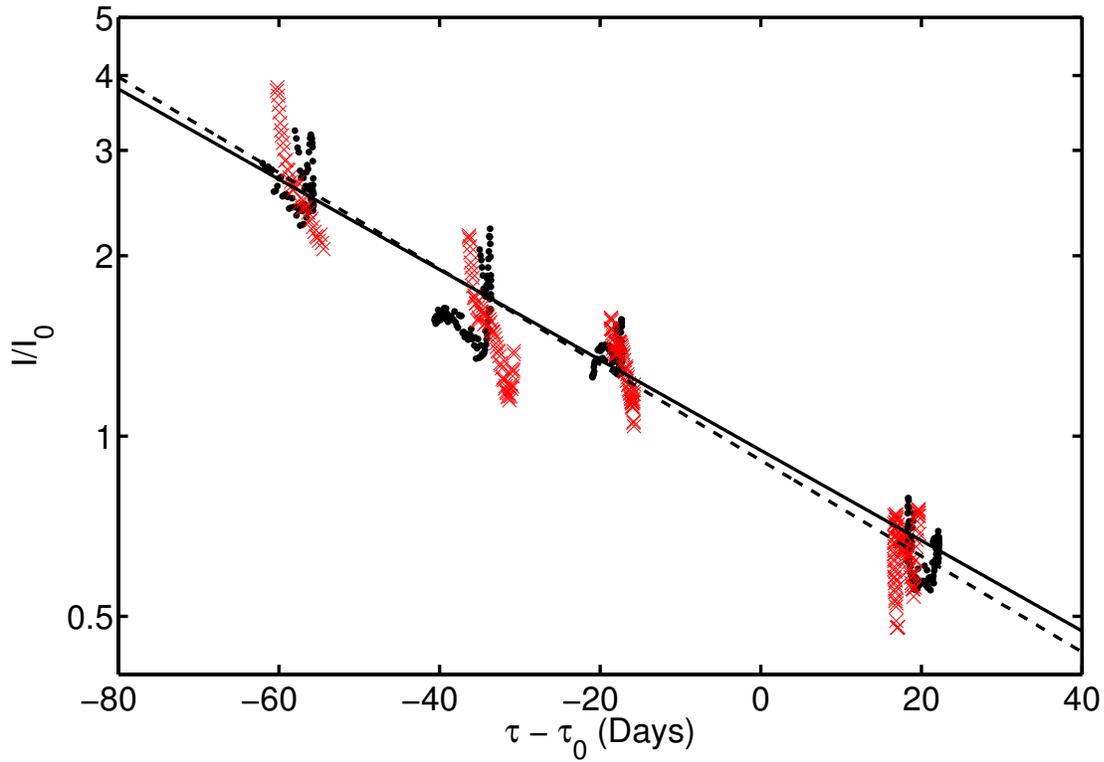}
	\caption[Brightness v. $\tau$ relation assuming exponential form]{The natural logarithm of the normalized intrinsic brightness, $I(t_b,\tau)/I_0(t_b,\tau_0)$, as a function of the difference $(\tau-\tau_0)$.  The trend will be linear if the brightness depends exponentially on $\tau$.  The values measured from the east and west jet are shown as black dots and red x-marks, respectively.  The best exponential fit is shown as the solid line for the east jet and the dashed line for the west jet. For the east jet $\tau'=57.6\pm1.4$ days, and for the west jet $\tau'=54.3\pm1.9$ days.}
	\label{fig:ivtau_exp}
\end{figure}

Figure \ref{fig:ivtau_exp} shows the ratio $I(t_{b},\tau)/I_{0}(t_{b},\tau_{0})$ as a function of $\tau - \tau_{0}$. On this plot an exponential will appear linear.  The data have been selected using the same criteria as in Figure \ref{fig:ivtau_pow}.  The best fit of equation \ref{eq:i_i0_tau_tau0_exp} to the east jet, shown as a solid line in the figure, gives $\tau'=57.6\pm1.4$ days.  The best fit to the west jet, shown as a dashed line in the figure, gives $\tau'=54.3\pm1.9$ days.  The exponential fits are better than the power law fits for both jets; the R$^2$ values for the fits are 0.93 and 0.90 for east and west, respectively. The fits are consistent with the east and west jets having the same value of $\tau'$, indicating that the external environments surrounding each jet are comparable. 

We note that there is some vertical scatter in each of these figures, and especially in Figure \ref{fig:ivtau_pow}.  This is not due to noise, rather it is the result of one of several possible systematic effects.  For example, if the assumed functional forms for $f(\tau)$ are not valid then this kind of scatter will be expected.  In Figure \ref{fig:ivtau_pow} we suspect that a large portion of the scatter is due to the inadequacy of the power law model.  Another possibility is that this is due to a breakdown in the assumption made in equation \ref{eq:I_is_factorable}, e.g. if $\tau_{0}$ varied slightly with $t_b$.  Nevertheless, from these figures we find that these kinds of systematic effects are secondary to the general, simple trends analyzed above. 

We conclude that the evolution of jet brightness in $\tau$, within the range $50 \lesssim \tau \lesssim 300$ days for both the east and west jets, is well described by an exponential function with $\tau'=55.9\pm1.7$ days.  This is a significantly longer exponential time constant than the value of $13-21$ days reported by \citet{jowett95}.  We note, however, that they were measuring the brightness evolution as a function of $t_{age}$ rather than $\tau$ and thus have not taken into account the varying amounts of light travel time from component to component.

\subsection{History of the Core Flux Variability}\label{ssec:core_variability}

\begin{figure}
	\centering
	\plotone{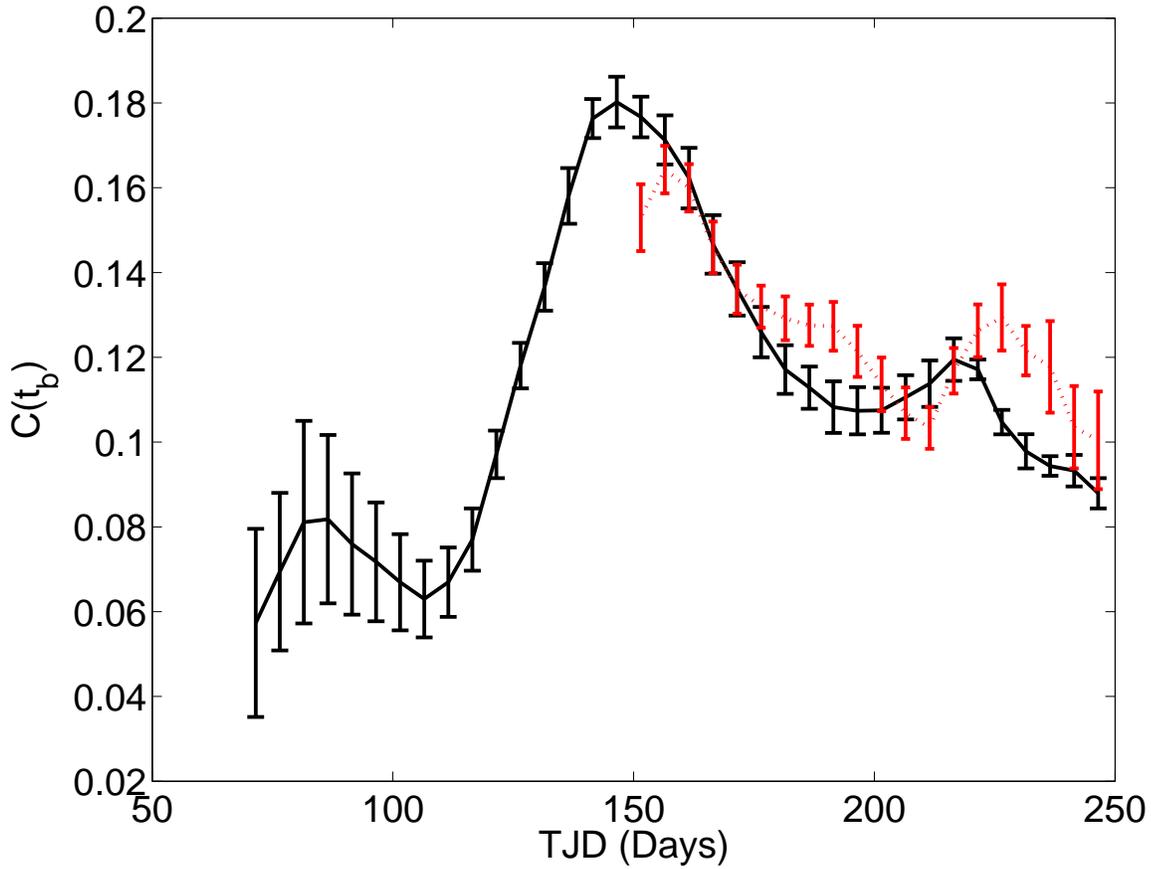}
	\caption[History of Core Variability]{The birth epoch dependent part of the intrinsic brightness, $C(t_b)$, as a function of birth epoch measured in TJD.  We compute $C$ assuming that the $\tau$ dependent part is an exponential,  $I(t_b, \tau) = C(t_b) e^{-\tau/\tau'}$, with $\tau' = 55.9$ days. The values derived for the east and west jets are shown in black and red, respectively.}
	\label{fig:brightness_coeff}
\end{figure}

If $f(\tau)$ has been determined then we can recover the proportionality constant $C$ for a given birth epoch simply:
\begin{equation} 
	C(t_{b})=I(t_{b},\tau)/f(\tau).
\end{equation}  
Here we assume $f(\tau)=e^{-\tau/\tau'}$ and we use $\tau'=55.9$ days.  Figure \ref{fig:brightness_coeff} shows this brightness coefficient as a function of birth epoch for both the east (shown in black) and west (shown in red) jets. This includes data from all five epochs.  The values measured for the west jet cover a much smaller range in $t_b$ because we have to exclude data measured from the ``loop''.  

The coefficient increases by more than a factor of five over the course of 50 days between TJD~100 and 150.  Where we have measured $C$ in both jets, the results are consistent with the east and west jet values being the same.      

\begin{figure}
	\centering
	\plotone{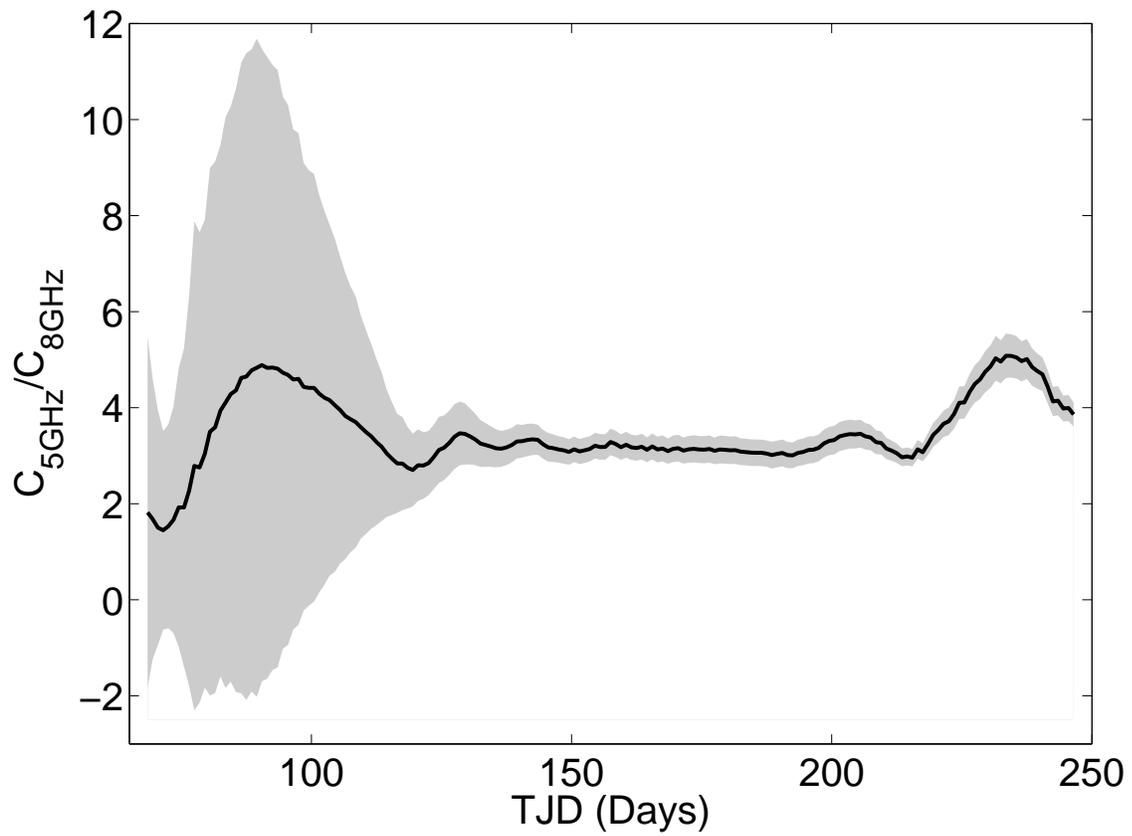}
	\caption[Ratio of C and X band core variability]{The ratio of $C(t_b)$ as measured from the east jet in the 5 and 8.5 GHz images. The shaded region indicates the uncertainty.}
	\label{fig:cx_ctb_ratio}
\end{figure}

The general trend is consistent between 5 and 8.5 GHz as well.  In Figure \ref{fig:cx_ctb_ratio} we show the ratio of the east jet coefficients measured from the 5 GHz profiles to those measured from the 8.5 GHz profiles ($C_{5GHz}(t_b)/C_{8.5GHz}(t_b)$).  The ratio remains roughly constant throughout the range in $t_b$.  Also note that, after accounting for the different beam dimensions, this ratio is consistent with the core flux ratio shown in Figure \ref{fig:core_v_tjd}.  The agreement between 5 and 8.5 GHz coefficients indicates variability with constant spectral index.  

The measurement of $C(t_b)$ provides a history of the relative power injected in the jet and could be used as a post-diction of the history of the radio core variability over the year preceding the observations. This interpretation suggests that the core flux would have increased dramatically over the course of a few weeks. This kind of outburst is not unprecedented, with a similar flare being detected during a series of VLA observations of SS433 in the summer of 2003 (Roberts et al., in preparation). We are not aware of any monitoring campaigns that span the period from mid 2006 to mid 2007, so this post-diction is not verifiable. Application of this procedure to VLA data taken in collaboration with a radio monitoring campaign would be required for confirmation. 

\section{Discussion}\label{sec:discussion}


\subsection{Comparing the Brightness Evolution to Radio Source Models}\label{ssec:comparison_to_models}

Now that we have a measurement of the brightness decay rate, we can compare the observed rate with model predictions. The simplest model is that of a spherically expanding radio source proposed by \citet{vanderLaan66}. In the case of SS\,433 we can think of the jets as a series of spherical blobs \citep{vermeulen87}.  In the van del Laan model, $S_{\nu}\propto r^{-2p}$, where $p=1+2\alpha$ is the exponent of the assumed electron energy distribution and $r$ is the radius of the source. In the case of ballistic expansion, where $r\propto\tau$, and since here $p \simeq 2.4$, this model predicts that the brightness will decay as a power law with an exponent of $-4.8$, which is much faster than we measure. If instead the jet expands at a slower rate, with $r\propto\tau^{2/5}$, then the decay would be proportional to $\tau^{-1.9}$ which is comparable to the power law fits to the data. An exponential decay of the brightness in time would require that the radius of the blobs expands exponentially.  If the brightness decays with an exponential time constant of $55.9$ days and $p=2.4$ then the radius will increase with an exponential time constant of $268$ days.  

A model of a precessing conical outflow was presented in \citet{hjellming88} to describe the observed jet profile in SS\,433 and other precessing jet systems. In the optically thin limit they find that the contribution to the flux density from each segment of the jet, $dz$, is $dS_{\nu}(z)\propto z^{-m}dz$, where $z$ is the distance away from the core (along the local velocity vector) and $m=(7p-1)/(6+6\delta)$. The parameter $\delta$ is 1 if the jet expansion is confined or 0 if it expands freely. Since the speed of the jet is assumed to be constant, $z\propto\tau$, so this also provides a relationship between the flux per unit length and time.  For $p\simeq2.4$, this results in $S_{\nu}\propto\tau^{-2.6}$ or $\tau^{-1.3}$ for the free and confined expansions, respectively. Comparison with our power law fits would suggest that the jets expand at a rate between these, with the east jet expanding more freely than the west.  However, we do not believe that this model appropriately describes the jets of SS\,433 because of an apparent flaw in the assumed geometry of the model.  

One of the key assumptions of the model proposed by \citet{hjellming88} is that there are no gradients in the velocity along the jet locus. Therefore the jets do not expand longitudinally but only laterally (2-D expansion). This assumption makes sense for a straight jet. For a helical jet, however, even though the magnitude of velocity is nearly the same for all parts of the jet and does not change with distance from the core as best we can measure (see Section \ref{ssec:decceleration}), the direction of the velocity vector varies from location to location. As a result, one component of the jet does not follow its nearest neighbor. Instead their trajectories diverge and expansion along the jet helix is inevitable.

\subsection{Comparing the Images to the Kinematic Model}\label{ssec:decceleration}

We know that the jet material is traveling ballistically through an external medium that is composed of the disk wind and other material within the surrounding supernova remnant. The jets will interact with the external environment and decelerate due to the resulting ram pressure. The jets observed in our radio images however, do not appear to have decelerated measurably. This is made clear by comparing the images to the kinematic model which successfully predicts the location of the jet out to $4''$ from the core. The parameters that we use to generate this model (apart from the distance to the source) have been derived independently from optical emission line data. The emission lines originate from a region smaller than $\sim10^{15}$ cm \citep{davidson80} while the arcsecond scale radio maps reveal the jets at a scale of $10^{17}$ cm. The model curve is computed assuming constant velocity. 

\citet{stirling04} found evidence for deceleration of 0.02c per period by comparing the kinematic model locus with radio images made using VLA and MERLIN data. It should be noted, however, that they assume a distance of 4.8 kpc where we adopt the more recently verified distance of 5.5 kpc from \citet{blundell04}. Calculating the model positions using a shorter distance will of course result in a larger projected size of the jets and accounts for the discrepancy between model and image that they noted. Inclusion of a 0.02c per period deceleration in the model calculation results in a compressed appearance of the jets that we do not observe.  Moreover, the kinematic model computed with a constant velocity fits our images equally well throughout the summer of 2007.  If the jets were decelerating we would expect the model to systematically lead the observed helix at increasing angular distance from the core.
 
The excellent agreement between the radio images and model curves is indicative of the stability of the velocity across a broad range of length scales.  We can use the agreement between the kinematic model and our radio images to put limits on the deceleration of jet material and use this to estimate the ratio of mass density between the jet and surrounding medium.   For simplicity, we assume that the jet is composed of spherical blobs of plasma. Each blob has radius $r$. If we relate the ram pressure acting on the blobs to the acceleration we find the following relationship between the density of the internal and external media,
\begin{eqnarray}
	K\rho_{e}v^{2}(\pi r^{2}) & = & \rho_{j}(\frac{4}{3}\pi r^{3})\dot{v},\nonumber \\
	\frac{\rho_{j}}{\rho_{e}} & = & \frac{3v^{2}}{4r\dot{v}},
\end{eqnarray}
where $\rho_{e}$ is the external density, $\rho_{j}$ is the jet density, $v$ is the blob velocity relative to the external medium,\footnote{\citet{begelman80} find that the disk wind travels with a velocity of 1500 km/s, which is only $\sim2\%$ of the jet speed ($\sim8\times10^{5}$ km/s). Thus we assume that the external environment is stationary relative to the jets.} and $K$ is a shape factor of order unity (we set $K=1$). The density $\rho_{j}$ will certainly decrease as the jets cool and expand. The density of the surrounding medium, $\rho_{e}$, will also likely decrease with increased distance from the binary system. For simplicity we assume that the densities decrease in such a way that their ratio is constant. 

Any acceleration would result in a displacement from the un-accelerated position (represented by the model curve) at a given time. The maximum displacement is $\Delta d\lesssim\frac{1}{2}\dot{v}t^{2}$. We will assume that this displacement projects to an angular distance that is less than the spatial resolution of our images since otherwise we would be able to detect it. An image resolution of $0.3$ arcseconds corresponds to a linear size of $2\times10^{16}$ cm, assuming a distance to SS\,433 of $5.5\mbox{ kpc}$. Substituting the maximum displacement for the acceleration we find 
\begin{equation}
	\frac{\rho_{j}}{\rho_{e}}=\frac{3(vt)^{2}}{8r\Delta d}.
\end{equation}
We measure the jets clearly out to an age of $\sim350$ days. For a ballistically expanding jet with a cone half-opening angle of $0.6^{\circ}$ \citep{marshall02} this ratio is $\rho_{j}/\rho_{e}\gtrsim300$. For a slower expansion rate the ratio will be larger.


\section{Conclusions}\label{sec:conclusions}

We have presented analysis of a sequence of five deep observations of SS\,433 made over the summer of 2007 using the VLA at 5 and 8 GHz. The main results are as follows.
\begin{enumerate}
\item We measure the spectral index to be $0.74\pm0.06$, consistent with previously reported values, and find no significant variations either across the source or with precession phase.   
\item All the results of Paper~II (based on a single very deep 5 GHz observation made in 2003) are confirmed, but with the sequence of five closely spaced observations, the possibilities for analysis are greatly increased. 
\item The jet intrinsic brightness profiles in 2007 are significantly different from those in 2003, even though the precession phases are similar. We attribute this to variability at the core, injecting varying power at the base of the jets.
\item As in Paper~II, in every image the profiles of the east and west jets are remarkably similar if projection and Doppler beaming are taken into account.
\item The sequence of five images allows us to disentangle the evolution of individual pieces of jet from variations of jet power originating at the core. We find that the brightness of individual pieces of jet fades as an exponential function of age (or distance from the core) rather than a power law. The exponential time constant is $55.9 \pm 1.7$ days. Our results are consistent with the brightness of the two jets evolving in the same way. 
\item There is also significant variation (by a factor of at least five) in jet power with birth epoch, with the east and west jets varying in synchrony. The variation is also consistent between 5 and 8.5 GHz.   
\item The lack of deceleration between the scale of the optical Balmer line emission ($10^{15}$ cm) and that of the radio emission ($10^{17}$ cm) requires that the jet material is much denser than its surroundings. We find that the density ratio must exceed 300:1.
\end{enumerate}


\appendix{}



\acknowledgements{}

This material is based upon work supported by the National Science Foundation under Grants Nos.~0307531 and 0607453 and prior grants. Any opinions, findings, and conclusions or recommendations expressed in this material are those of the authors and do not necessarily reflect  the views of the National Science Foundation. D.H.R.\  gratefully acknowledges the support of the William R.\ Kenan, Jr.\ Charitable Trust. The National Radio Astronomy Observatory is a facility of the National Science Foundation, operated under cooperative agreement by Associated Universities, Inc.  For plotting and image analysis, we have made use of the CFITSIO and MFITSIO packages made available by the NASA Goddard SFC and Damian Eads of LANL, respectively.

Facilities: \facility{VLA/EVLA (A array)}



\begin{deluxetable}{ccccc}
	\tablecaption{Observational Summary\label{tab:obs_overview}}
	
	\tablecolumns{5}
	
	\tablewidth{4.5in}
	
	\tablehead{\colhead{} & \colhead{} & \multicolumn{2}{c}{\scriptsize{\# Antennas (VLA/EVLA)}} & \colhead{} \\ \colhead{Date} & \colhead{TJD\tablenotemark{b}} & \colhead{C-Band} & \colhead{X-Band} & \colhead{Precessional Phase\tablenotemark{a}} } 

	\startdata
		2008 June 6  &  260 &  16/4 &  16/9 &  0.72 \\
		2007 July 1  &  283 &  17/5 &  17/8 &  0.86 \\
		2007 July 18 &  300 &  16/6 &  15/9 &  0.96 \\
		2007 Aug. 5  &  318 &  14/7 &  16/9 &  0.07 \\
		2007 Aug. 24 &  337 &  15/5 &  16/7 &  0.19 \\
	\enddata

	\tablenotetext{a}{Phase is zero when the east jet is maximally redshifted following \citet{vermeulen93}.}	
	\tablenotetext{b}{Truncated Julian Date in days, TJD = JD - 2454000}

\end{deluxetable}

\begin{deluxetable}{ccccccc}

	\tablecaption{C-band Image Details\label{tab:c_image_details}}
	
	\tablecolumns{7}
	
	\tablewidth{5.5in}
	
	\tablehead{\colhead{Date} & \colhead{BMAJ} & \colhead{BMIN} & \colhead{BPA} & \colhead{$\mathrm{I}_{max}$} & \colhead{$\mathrm{I}_{min}$} & \colhead{RMS} \\ 
	\colhead{} & \colhead{arcsec} & \colhead{arcsec} & \colhead{\small{Degrees}} & \colhead{\small{mJy/Beam}} & \colhead{\small{mJy/Beam}} & \colhead{\small{mJy/Beam}} } 

	\startdata
		2007 June 8  &  0.404 &  0.340 &  $-68$ & 329 &  0.12 & 0.04 \\
		2007 July 1  &  0.381 &  0.335 &  $-22$ & 359 &  0.10 & 0.03 \\
		2007 July 18 &  0.410 &  0.348 &  $-6$9 & 261 &  0.11 & 0.04 \\
		2007 Aug. 5  &  0.414 &  0.345 &  $-88$ & 225 &  0.14 & 0.05 \\
		2007 Aug. 24 &  0.356 &  0.330 &  \phantom{$-$}$35$  & 467 &  0.19 & 0.05 \\
	\enddata

\end{deluxetable}

\begin{deluxetable}{ccccccc}

	\tablecaption{X-band Image Details\label{tab:x_image_details}}
	
	\tablecolumns{7}
	
	\tablewidth{5.5in}
	
	\tablehead{\colhead{Date} & \colhead{BMAJ} & \colhead{BMIN} & \colhead{BPA} & \colhead{$\mathrm{I}_{max}$} & \colhead{$\mathrm{I}_{min}$} & \colhead{RMS} \\ 
	\colhead{} & \colhead{arcsec} & \colhead{arcsec} & \colhead{\small{Degrees}} & \colhead{\small{mJy/Beam}} & \colhead{\small{mJy/Beam}} & \colhead{\small{mJy/Beam}} } 

	\startdata
		2007 June 8  &  0.219 & 0.208 & \phantom{$-$}$53$  & 211 & 0.08 & 0.03 \\ 
		2007 July 1  &  0.219 & 0.201 & $-31$ & 221 & 0.08 & 0.03 \\
		2007 July 18 &  0.224 & 0.209 & \phantom{$-$}$63$  & 141 & 0.09 & 0.03 \\
		2007 Aug. 5  &  0.238 & 0.205 & $-73$ & 129 & 0.07 & 0.02 \\
		2007 Aug. 24 &  0.240 & 0.213 & $-49$ & 361 & 0.11 & 0.04 \\
	\enddata

\end{deluxetable}

\begin{deluxetable}{cccccc}

	\tablecaption{Profile Plot Reference Times ($\tau$ in days)\label{tab:reference_times}}
	
	\tablecolumns{6}
	
	
	\tablehead{\colhead{Obs. Date} & \colhead{$\theta_e = \theta_w = 90^\circ$} & \colhead{$\theta_e = \mathrm{min.}$} & \colhead{$\theta_e = \mathrm{max.}$} & \colhead{$\theta_w = \mathrm{min.}$} & \colhead{$\theta_w = \mathrm{max.}$} }

	\startdata
		2007 June 8  & 91, 142, 253, 304      & 41, 231, 420 & 110, 267     & 121, 291     & 30, 173, 315 \\
		2007 July 1  & 114, 165, 276, 327     & 67, 257, 447 & 133, 289     & 146, 315     & 52, 192, 335 \\
		2007 July 18 & 20, 131, 182, 293, 344 & 88, 278      & 150, 305     & 163, 333     & 66, 208, 350 \\
		2007 Aug. 5  & 38, 149, 200, 311, 362 & 107,298      & 12, 168, 333 & 12, 182, 352 & 81, 223 \\
		2207 Aug. 24 & 57, 168, 219, 330, 380 & 131, 320     & 30, 185, 341 & 32, 202, 370 & 98, 240 \\
	\enddata

\end{deluxetable}

%
%
%
%
%
%


\bibliographystyle{apj}
\bibliography{ss433_refs}

\begin{thebibliography}{30}
\expandafter\ifx\csname natexlab\endcsname\relax\def\natexlab#1{#1}\fi

\bibitem[{{Begelman} {et~al.}(1980){Begelman}, {Hatchett}, {McKee}, {Sarazin},
  \& {Arons}}]{begelman80}
{Begelman}, M.~C., {Hatchett}, S.~P., {McKee}, C.~F., {Sarazin}, C.~L., \&
  {Arons}, J. 1980, \apj, 238, 722

\bibitem[{Bell(2010)}]{bell_thesis}
Bell, M. 2010, PhD thesis, Brandeis University

\bibitem[{Blandford \& Konigl(1979)}]{blandford79}
Blandford, R.~D., \& Konigl, A. 1979, \apj, 232, 34

\bibitem[{Blundell \& Bowler(2004)}]{blundell04}
Blundell, K.~M., \& Bowler, M.~G. 2004, \apj, 616, L159

\bibitem[{{Blundell} {et~al.}(2007){Blundell}, {Bowler}, \&
  {Schmidtobreick}}]{blundell07}
{Blundell}, K.~M., {Bowler}, M.~G., \& {Schmidtobreick}, L. 2007, \aap, 474,
  903

\bibitem[{{Davidson} \& {McCray}(1980)}]{davidson80}
{Davidson}, K., \& {McCray}, R. 1980, \apj, 241, 1082

\bibitem[{{De Young}(2002)}]{deyoung02}
{De Young}, D.~S. 2002, The Physics of Extragalactic Radio Sources (The
  University of Chicago Press)

\bibitem[{Dubner {et~al.}(1998)Dubner, Holdaway, Goss, \& Mirabel}]{dubner98}
Dubner, G.~M., Holdaway, M., Goss, W.~M., \& Mirabel, I.~F. 1998, \aj, 116,
  1842

\bibitem[{Eikenberry {et~al.}(2001)Eikenberry, Cameron, Fierce, Kull, Dror,
  Houck, \& Margon}]{eikenberry01}
Eikenberry, S.~S., Cameron, P.~B., Fierce, B.~W., Kull, D.~M., Dror, D.~H.,
  Houck, J.~R., \& Margon, B. 2001, \apj, 561, 1027

\bibitem[{{Elston} \& {Baum}(1987)}]{elston87}
{Elston}, R., \& {Baum}, S. 1987, \aj, 94, 1633

\bibitem[{Fabrika(2004)}]{fabrika06}
Fabrika, S. 2004, Astrophysics and Space Physics Reviews, 12, 1

\bibitem[{Hjellming \& Johnston(1981)}]{hjellming81}
Hjellming, R.~M., \& Johnston, K.~J. 1981, \apj, 246, L141

\bibitem[{{Hjellming} \& {Johnston}(1988)}]{hjellming88}
{Hjellming}, R.~M., \& {Johnston}, K.~J. 1988, \apj, 328, 600

\bibitem[{{Jowett} \& {Spencer}(1995)}]{jowett95}
{Jowett}, F.~H., \& {Spencer}, R.~E. 1995, in The XXVIIth Young European Radio
  Astronomers Conference, ed. {D.~A.~Green \& W.~Steffen}, 12

\bibitem[{{Katz} {et~al.}(1982){Katz}, {Anderson}, {Grandi}, \&
  {Margon}}]{katz82}
{Katz}, J.~I., {Anderson}, S.~F., {Grandi}, S.~A., \& {Margon}, B. 1982, \apj,
  260, 780

\bibitem[{Margon(1984)}]{margon84}
Margon, B. 1984, \araa, 22, 507

\bibitem[{{Margon} {et~al.}(1979{\natexlab{a}}){Margon}, {Ford}, {Katz},
  {Kwitter}, {Ulrich}, {Stone}, \& {Klemola}}]{margon79a}
{Margon}, B., {Ford}, H.~C., {Katz}, J.~I., {Kwitter}, K.~B., {Ulrich}, R.~K.,
  {Stone}, R.~P.~S., \& {Klemola}, A. 1979{\natexlab{a}}, \apjl, 230, L41

\bibitem[{{Margon} {et~al.}(1979{\natexlab{b}}){Margon}, {Grandi}, {Stone}, \&
  {Ford}}]{margon79b}
{Margon}, B., {Grandi}, S.~A., {Stone}, R.~P.~S., \& {Ford}, H.~C.
  1979{\natexlab{b}}, \apjl, 233, L63

\bibitem[{Marshall {et~al.}(2002)Marshall, Canizares, \& Schulz}]{marshall02}
Marshall, H.~L., Canizares, C.~R., \& Schulz, N.~S. 2002, \apj, 564, 941

\bibitem[{{Mirabel} \& {Rodr{\'{\i}}guez}(1999)}]{mirabel99}
{Mirabel}, I.~F., \& {Rodr{\'{\i}}guez}, L.~F. 1999, \araa, 37, 409

\bibitem[{NRAO(2004)}]{AIPS}
NRAO. 2004, AIPS Cookbook, NRAO, Socorro, NM

\bibitem[{Paragi {et~al.}(1999)Paragi, Vermeulen, Fejes, Schilizzi, Spencer, \&
  Stirling}]{paragi99}
Paragi, Z., Vermeulen, R.~C., Fejes, I., Schilizzi, R.~T., Spencer, R.~E., \&
  Stirling, A.~M. 1999, \aap, 348, 910

\bibitem[{{Roberts} {et~al.}(2010){Roberts}, {Wardle}, {Bell}, {Mallory},
  {Marchenko}, \& {Sanderbeck}}]{roberts10}
{Roberts}, D.~H., {Wardle}, J.~F.~C., {Bell}, M.~R., {Mallory}, M.~R.,
  {Marchenko}, V.~V., \& {Sanderbeck}, P.~U. 2010, \apj, 719, 1918

\bibitem[{Roberts {et~al.}(2008)Roberts, Wardle, Lipnick, Selesnick, \&
  Slutsky}]{roberts08}
Roberts, D.~H., Wardle, J. F.~C., Lipnick, S.~L., Selesnick, P.~L., \& Slutsky,
  S. 2008, \apj, 676, 584

\bibitem[{{Seaquist} {et~al.}(1980){Seaquist}, {Gilmore}, {Nelson}, {Payten},
  \& {Slee}}]{1980ApJ...241L..77S}
{Seaquist}, E.~R., {Gilmore}, W., {Nelson}, G.~J., {Payten}, W.~J., \& {Slee},
  O.~B. 1980, \apjl, 241, L77

\bibitem[{{Shepherd} {et~al.}(1994){Shepherd}, {Pearson}, \&
  {Taylor}}]{shepherd94}
{Shepherd}, M.~C., {Pearson}, T.~J., \& {Taylor}, G.~B. 1994, in Bulletin of
  the American Astronomical Society, Vol.~26, Bulletin of the American
  Astronomical Society, 987--989

\bibitem[{Stirling {et~al.}(2004)Stirling, Spencer, Cawthorne, \&
  Paragi}]{stirling04}
Stirling, A.~M., Spencer, R.~E., Cawthorne, T.~V., \& Paragi, Z. 2004, \mnras,
  354, 1239

\bibitem[{{van der Laan}(1966)}]{vanderLaan66}
{van der Laan}, H. 1966, \nat, 211, 1131

\bibitem[{Vermeulen {et~al.}(1987)Vermeulen, Icke, Schilizzi, Fejes, \&
  Spencer}]{vermeulen87}
Vermeulen, R.~C., Icke, V., Schilizzi, R.~T., Fejes, I., \& Spencer, R.~E.
  1987, \nat, 328, 309

\bibitem[{{Vermeulen} {et~al.}(1993){Vermeulen}, {Schilizzi}, {Spencer},
  {Romney}, \& {Fejes}}]{vermeulen93}
{Vermeulen}, R.~C., {Schilizzi}, R.~T., {Spencer}, R.~E., {Romney}, J.~D., \&
  {Fejes}, I. 1993, \aap, 270, 177

\end{thebibliography}

\onecolumn

\end{document}